\documentclass[
twocolumn,
]{aastex631}
\usepackage{subfigure}
\usepackage{amsmath}
\usepackage{txfonts}
\usepackage[utf8]{inputenc}

\shorttitle{Impact of Rastall Gravity on Mass, Radius and Sound Speed of the Pulsar PSR J0740+6620}
\shortauthors{Waleed El Hanafy}
\graphicspath{{./}{figures/}}
\begin{document}

\title{Impact of Rastall Gravity on Mass, Radius and Sound Speed of the Pulsar PSR J0740+6620}

\email{waleed.elhanafy@bue.edu.eg}

\author[0000-0002-0097-6412]{Waleed El Hanafy}
\affiliation{Centre for Theoretical Physics, The British University in Egypt, P.O. Box 43, El Sherouk City, Cairo 11837, Egypt}

%% Note that the \and command from previous versions of AASTeX is now
%% depreciated in this version as it is no longer necessary. AASTeX
%% automatically takes care of all commas and "and"s between authors names.

%% AASTeX 6.31 has the new \collaboration and \nocollaboration commands to
%% provide the collaboration status of a group of authors. These commands
%% can be used either before or after the list of corresponding authors. The
%% argument for \collaboration is the collaboration identifier. Authors are
%% encouraged to surround collaboration identifiers with ()s. The
%% \nocollaboration command takes no argument and exists to indicate that
%% the nearby authors are not part of surrounding collaborations.

%% Mark off the abstract in the ``abstract'' environment.
\begin{abstract}
Millisecond pulsars are perfect laboratories to test possible matter-geometry coupling and its physical implications in light of recent Neutron Star Interior Composition Explorer (NICER) observations. We apply Rastall field equations of gravity, where matter and geometry are nonminimally coupled, to Krori-Barua interior spacetime whereas the matter source is assumed to be anisotropic fluid. We show that all physical quantities inside the star can be expressed in terms of Rastall, $\epsilon$, and compactness, $C=2GM/Rc^2$, parameters. Using NICER and X-ray Multi-Mirror X-ray observational constraints on the mass and radius of the pulsar PSR J0740+6620 we determine Rastall parameter to be at most $\epsilon=0.041$ in the positive range. The obtained solution provides a stable compact object; in addition the squared sound speed does not violate the conjectured sound speed $c_s^2\leq c^2/3$ unlike the general relativistic treatment. We note that no equations of state are assumed; the model however fits well with linear patterns with bag constants. In general, for $\epsilon>0$, the theory predicts a slightly larger size star in comparison to general relativity for the same mass. This has been explained as an additional force, due to matter-geometry coupling, in the hydrodynamic equilibrium equation, which contributes to partially diminish the gravitational force effect. Consequently, we calculate the maximal compactness as allowed by the strong energy condition to be $C = 0.735$ which is $\sim 2\%$ higher than general relativity prediction. Moreover, for the surface density at saturation nuclear density $\rho_{\text{nuc}} = 2.7\times 10^{14}$ g/cm$^3$ we estimate the maximum mass $M=4 M_\odot$ at radius $R=16$ km.
\end{abstract}

%% Keywords should appear after the \end{abstract} command.
%% The AAS Journals now uses Unified Astronomy Thesaurus concepts:
%% https://astrothesaurus.org
%% You will be asked to selected these concepts during the submission process
%% but this old "keyword" functionality is maintained in case authors want
%% to include these concepts in their preprints.
%\keywords{Classical Novae (251) --- Ultraviolet astronomy(1736) --- History of astronomy(1868) --- Interdisciplinary astronomy(804)}
\keywords{Massive stars (732) --- Millisecond pulsars (1062) --- Neutron star cores (1107) --- Non-standard theories of gravity (1118) --- Stellar structures (1631)}

%% From the front matter, we move on to the body of the paper.
%% Sections are demarcated by \section and \subsection, respectively.
%% Observe the use of the LaTeX \label
%% command after the \subsection to give a symbolic KEY to the
%% subsection for cross-referencing in a \ref command.
%% You can use LaTeX's \ref and \label commands to keep track of
%% cross-references to sections, equations, tables, and figures.
%% That way, if you change the order of any elements, LaTeX will
%% automatically renumber them.
%%
%% We recommend that authors also use the natbib \citep
%% and \citet commands to identify citations.  The citations are
%% tied to the reference list via symbolic KEYs. The KEY corresponds
%% to the KEY in the \bibitem in the reference list below.

\section{Introduction}\label{Sec:Introduction}

Neutron stars (NSs) provide a unique laboratory to test matter at extremely high density conditions, a few times the nuclear saturation density $\rho_\text{nuc} \approx 2.7 \times 10^{14}$ g/cm$^3$, which are terrestrially inaccessible. The nature of their cores are poorly understood due to lacking of knowledge of equation of state (EoS) at these densities \citep{Ozel:2016oaf}. Many EoSs are proposed to describe the neutron star cores by including baryonic or exotic matter or possible combinations. One way to test these EoSs is to simultaneously determine the NS mass and radius. This represents the primary aim of several astrophysical observations nowadays. The pulsar timing measurement--which depends on the determination of radio pulse times of arrival in comparison with a stable reference clock--of pulses is being used for many decades to determine the pulsars' masses. The pulse period for millisecond pulsars (MSPs) ranges from 1.39 to 30 ms, slow down rate $ \leq 10^{-19}$ s/s with a characteristic age $10^9$ yrs. Those are very useful to test relativistic theories \citep[c.f.,][]{Stairs:2003eg,Reardon:2015kba}.

Recently X-ray and gravitational wave signal observations have opened new windows to determine NSs masses and radii. NICER mission is devoted to measure mass and radius (and consequently EoS) of MSPs by analyzing the gravitational light bending and the X–ray light curves produced by rotating hot spots on their surfaces \citep{Bogdanov:2019ixe,Bogdanov:2019qjb}. Some MSP with masses $\sim 1.5 M_\odot$ have been observed and analyzed by NICER, e.g. PSR J0030+0451 with mass $M=1.44^{+0.15}_{-0.14} M_\odot$ \citep{Miller:2019cac} and independently by \citet{Raaijmakers:2019qny} where $M= 1.34^{+0.15}_{-.16} M_\odot$, while others, e.g. PSR J0437-4715 with mass $M=1.44 \pm 0.07 M_\odot$ \citep{Reardon:2015kba}, are in progress \citep{2014HEAD...1411607B}. Notably, for the isolated pulsar PSR J0030+0451, two independent analyses of NICER data have been done to measure its radius, $R= 13.02_{-1.06}^{+1.24}$ km \citep{Miller:2019cac} and $R= 12.71^{+1.14}_{-1.19}$ km \citep{Raaijmakers:2019qny}. Although these studies assume different patterns of the hot spots, both agree on the measured value of the pulsar radius. Additionally, some MSP with mass $\sim 2 M_\odot$, e.g. PSR J1614-2230 with mass $M= 1.908 \pm 0.016 M_\odot$ \citep{Demorest:2010bx,Fonseca:2016tux,NANOGRAV:2018hou}, PSR J0348+0432 with mass $M= 2.01 \pm 0.04 M_\odot$ \citep{Antoniadis:2013pzd} and PSR J0740+6620 with mass $M= 2.08 \pm 0.07 M_\odot$ using the relativistic Shapiro time delay \citep{NANOGrav:2019jur,Fonseca:2021wxt}, are of a great interest since they approach the upper limit of a NS or the lower limit of a Black Hole (BH). In particular, the pulsar PSR J0740+6020 is in a binary system which provides an independent measurements of mass and inclination. So its mass can be determined with higher precision, however its low NICER count rate relative to PSR J0348+0432 represents a challenge. Notably both NICER and X-ray Multi-Mirror (XMM) Newton data sets have been used to determine the radius of the pulsar PSR J0740+6020 by \cite{Miller:2021qha}, $R= 13.7_{-1.5}^{+2.6}$ km, and independently by \cite{Riley:2021pdl}, $R=12.39_{-0.98}^{+1.30}$ km (68\% credible level), while from NICER+XMM based on Gaussian process applying a nonparameteric EoS approach, the mass and radius are given as $M=2.07 \pm 0.11 M_\odot$ and $R=12.34^{+1.89}_{-1.67}$ km \citep{Legred:2021hdx} which is in agreement with \cite{Landry:2020vaw} at 68\% credible level.

Gravitational wave signals provide another window to test matter at high density. This research is led by Laser Interferometer Gravitational-Wave Observatory (LIGO) and Virgo collaboration \citep{Abbott:2016blz}. For example, GW190814, a binary compact system merger with a companion of mass $\sim 2.6 M_\odot$ \citep{LIGOScientific:2020zkf}, provides the first observed compact object in the hypothesized lower mass gap 2.5--5 $M_\odot$. This may point out some kind of modified gravity as suggested by \cite{Moffat:2020jic}. Also, GW170817 \citep{TheLIGOScientific:2017qsa,LIGOScientific:2018cki} and GW190425 \citep{LIGOScientific:2020aai} signals indicate no tidal deformation in the observed patterns of their gravitational waves, which in turn inform that the EoS is not too stiff. This is unlike what is usually assumed for high mass compact objects.

Remarkably, NICER observations of PSR J0030+0451 and PSR J0740+6020 provide an evidence against more squeezable models. The latter has much more mass than the former while both are having almost same size. Then it is reasonable to assume some mechanisms to justify the non-squeezability of a NS when it gains more mass. On the other hand, existence of high-mass pulsars $\sim 2 M_\odot$ such as PSR J0740+6020 is known to favor violation of the upper sound speed conformal limit $c_s^2= c^2/3$ which represents another challenge for theoretical models even at low density cases as shown by \cite{Bedaque:2014sqa} \citep[see also][]{Cherman:2009tw,Landry:2020vaw}. For the pulsar PSR J0740+6020, in particular, \cite{Legred:2021hdx} in their study concluded that the conformal sound speed is strongly violated at the NS core whereas $c_s^2=0.75 c^2$ with density $\sim 3.60\, \rho_\text{nuc}$.

When a NS gains more mass, one would expect stronger gravity and then the NS collapses to smaller size (consequently higher density). However, the NICER observations of the pulsars PSR J0030+0451 and PSR J0740+6020 do not agree with this mechanism. We argue that at high mass NS, one expect the presence of high density to form anisotropy \citep{herrera1997local}, whereas radial, $p_r$, and tangential, $p_t$, pressures are different. The induced anisotropic force becomes repulsive when $p_t>p_r$, subsequently acts against the attractive gravitational force as can be seen from Tolman–Oppenheimer–Volkoff (TOV) equation \citep{bowers1974anisotropic}. This in return supports the NS to hold its size. On the other hand, we suggest that a nonminimal coupling between matter and geometry inside compact objects (high curvatured spacetime) to play a crucial role to hold the conjectured sound speed limit from the NS core to its surface unlike the GR case.

In fact, the null covariant divergence of the energy-momentum tensor is one of the fundamental assumptions of the General Relativity (GR). On the contrary, Rastall attempted to modify GR by dropping this assumption assuming that the covariant derivative of the energy-momentum tensor to be proportional to Ricci scalar derivative, which directly reflects the nonminimal coupling between matter and geometry in curved spacetimes \cite{Rastall:1972swe,Rastall:1976uh}. In flat spacetimes the interaction term vanishes and GR would be recovered indeed. Rastall theory (RT) could be strongly tested within stellar structure models in the light of the recently unprecedented improvements of mass--radius observations of astrophysical objects as discussed above.

We arrange the present study as follows: In Sec. \ref{Sec:Rastall}, we review the modification of the general relativity as proposed by Rastall. In Sec. \ref{Sec:Model}, we setup the main assumptions of the present study. In Sec. \ref{Sec:Stability}, we use the X-ray NICER and XMM observations of the PSR J0740+6620 to constrain the model parameters. Also, we discuss the physical features of the pulsar and its stability as obtained by the present model. In Sec. \ref{Sec:pulsars}, we confront the model to more pulsars data. In Sec. \ref{Sec:MR-reln}, we estimate the maximum compactness as allowed by physical conditions. Consequently, we plot the mass-radius curves for different choices of surface densities showing the maximal mass in each case for a stable configuration. In Sec. \ref{Sec:Conclusion}, we conclude the work.
%%%%%%%%%%%%%%%%%%%%%%%%%%%%%%%%%%%%%%%%%%%%%% Section 2 %%%%%%%%%%%%%%%%%%%%%%%%%%%%%%%%%%%%%%%%%%%%%%%%%%%%%%%%%%%%%%%
\section{Rastall Gravity}\label{Sec:Rastall}
The contracted Bianchi identity, in Riemannian geometry, implies a covariant divergence-free Einstein tensor
\begin{equation}\label{Bianchi}
    \nabla{_\alpha}\mathfrak{G}_{\alpha\beta}=  \nabla{_\alpha}(\mathfrak{R}_{\alpha\beta}-\frac{1}{2}g_{\alpha\beta}\mathfrak{R})\equiv 0,
\end{equation}
where $\mathfrak{G}_{\alpha\beta}$ denotes Einstein tensor, $\mathfrak{R}_{\alpha\beta}$ denotes Ricci tensor and $\mathfrak{R}=g^{\alpha\beta}\mathfrak{R}_{\alpha\beta}$ denotes Ricci invariant. On the other hand, applying the minimal coupling procedure one requires a divergence-free energy-momentum
\begin{equation}\label{MCP}
    \nabla_{\alpha}\mathfrak{T}{^\alpha}{_\beta}=0.
\end{equation}
Therefore, the GR field equations, as formulated by Einstein, are given
\begin{equation}\label{GR}
    \mathfrak{G}_{\alpha\beta}=\kappa_E \mathfrak{T}_{\alpha\beta},
\end{equation}
where the Einstein coupling constant $\kappa_E=8\pi G/c^4$ with $G$ is the Newtonian gravitational constant and $c$ is the speed of light.

In Rastall's gravity the minimal coupling assumption \eqref{MCP} has been replaced by \citep{Rastall:1972swe,Rastall:1976uh}
\begin{equation}\label{Rastall}
     \nabla_{\alpha}\mathfrak{T}{^\alpha}{_\beta}=\frac{\epsilon}{\kappa}\, \partial_{\beta} \mathfrak{R}.
\end{equation}
Clearly the nondivergence-free energy-momentum assumption--proportional to Ricci scalar derivative--reflects a nonminimal coupling between matter and geometry, where $\epsilon/\kappa$ is a constant of the proportionality. The local violation of the conservation law, as assumed by Rastall, is a manifestation of an interaction between matter and curved spacetime, which no longer exists when spacetime is being flat. This, accordingly, led Rastall to obtain a some kind of generalized field equations
\begin{equation}\label{RT}
    \mathfrak{G}_{\alpha\beta}=\mathfrak{R}_{\alpha\beta}-\frac{1}{2}g_{\alpha\beta}\mathfrak{R}=\kappa(\mathfrak{T}_{\alpha\beta}-\frac{\epsilon}{\kappa} g_{\alpha\beta}\mathfrak{R}).
\end{equation}
For vacuum solutions, obviously, the GR theory is restored. By contracting Eq. \eqref{RT} we write
\begin{equation}\label{R-contraction}
    \mathfrak{R}=-\frac{\kappa}{1-4\epsilon} \mathfrak{T},\quad (\epsilon\neq \frac{1}{4})
\end{equation}
where $\mathfrak{T}=g^{\alpha\beta}\mathfrak{T}_{\alpha\beta}$ is the trace of the energy-momentum tensor. Alternatively, the field equations \eqref{RT} can be written as
\begin{equation}\label{RT2}
     \mathfrak{G}_{\alpha\beta}=\kappa \widetilde{\mathfrak{T}}_{\alpha\beta},
\end{equation}
where the effective energy-momentum tensor is given by
\begin{equation}\label{RTmn}
    \widetilde{\mathfrak{T}}_{\alpha\beta}=\mathfrak{T}_{\alpha\beta}+\frac{\epsilon}{1-4\epsilon}g_{\alpha\beta}\mathfrak{T}. \qquad (\epsilon\neq \frac{1}{4})
\end{equation}
The Newtonian limit of Rastall gravity requires a rescaled effective gravitational constant \citep[c.f.][]{Rastall:1972swe,Moradpour:2017ycq,Moradpour:2016ubd}
\begin{equation}\label{Rconst}
    \kappa=\eta\,\kappa_E \text{ and } \eta=\frac{1-4\epsilon}{1-6\epsilon}, \qquad (\epsilon\neq \frac{1}{6})
\end{equation}
where $\kappa$ acquires the dimension of $\kappa_E$ and $\kappa=\kappa_E$ only if $\epsilon=0$ (i.e. $\eta=1$). In practice, the deviations from GR are given in terms of the dimensionless Rastall parameter $\epsilon$ via the effective energy-momentum tensor $\widetilde{\mathfrak{T}}_{\alpha\beta}$ as well as the effective gravitational constant $\kappa$. For $\epsilon=0$ the conservation law is restored and the GR version of gravity is recovered. In other words, RT generalizes GR by dropping the divergence-free energy-momentum assumption and replaces it by assuming that the conservation law can be violated in curved spacetime due to existence of a nonminimal coupling between matter and geometry. It has been argued that RT is a reproduction of GR and both theories are completely equivalent as noted by \citet{Visser:2017gpz}. On the contrary, \citet{Darabi:2017coc} have investigated that claim showing that the matter-geometry coupling term in RT has been misinterpreted which led finally to a wrong conclusion. A counter example has been given by applying Visser's argument to $f(R)$ gravity, this led to an equivalence between $f(R)$ and GR as well which is not true. In this sense, stellar structure models are important applications to examine RT whereas the presence of matter (consequently curved spacetime) plays a crucial role to determine matter-geometry coupling, if any.
%%%%%%%%%%%%%%%%%%%%%%%%%%%%%%%%%%%%%%%%%%%%%% Section 3 %%%%%%%%%%%%%%%%%%%%%%%%%%%%%%%%%%%%%%%%%%%%%%%%%%%%%%%%%%%%%%%
\section{The Model}\label{Sec:Model}
We consider a static spherically symmetric line element for a 4-dimensional spacetime which can be expressed in a spherical polar coordinates ($t,r,\theta,\phi$)
\begin{equation}\label{eq:metric}
    ds^2=-e^{\alpha(r)}c^2 dt^2 + e^{\beta(r)} dr^2+ r^2 (d\theta^2+\sin^2 \theta \, d\phi^2),
\end{equation}
where $\alpha(r)$ and $\beta(r)$ be the metric potentials. We further assume the energy-momentum tensor for a anisotropic fluid with spherical symmetry, i.e.
\begin{equation}\label{Tmn-anisotropy}
    \mathfrak{T}{^\alpha}{_\beta}=(p_{t}+\rho c^2)U{^\alpha} U{_\beta}+p_{t} \delta ^\alpha _\beta + (p_{r}-p_{t}) V{^\alpha} V{_\beta},
\end{equation}
where $\rho=\rho(r)$ is the fluid energy density, $p_{r}=p_{r}(r)$ its radial pressure (in the direction the time-like four-velocity $U_\alpha$), $p_{t}=p_{t}(r)$ its tangential pressure (perpendicular to $U_\alpha$) and $V{^\alpha}$ is the unit space-like vector in the radial direction. Then, the energy-momentum tensor takes the diagonal form $\mathfrak{T}{^\alpha}{_\beta}=diag(-\rho c^2,\,p_{r},\,p_{t},\,p_{t})$.
Applying Rastall's field equations \eqref{RT2} to the spacetime \eqref{eq:metric} where the matter sector is as given by \eqref{Tmn-anisotropy}, we obtain the following
\begin{eqnarray}
\kappa \rho c^2&=&\frac{e^{-\beta}}{r^2}(e^\beta+\beta' r -1)-{\epsilon \kappa \over 1-4\epsilon}(\rho c^2-p_r-2p_t)\,,\nonumber\\
\kappa p_r&=&\frac{e^{-\beta}}{r^2}(1-e^\beta+\alpha' r)+{\epsilon \kappa \over 1-4\epsilon}(\rho c^2-p_r-2p_t)\,,\nonumber\\
\kappa p_t&=&\frac{e^{-\beta}}{4 r}\left[(2\alpha''-\alpha' \beta'+\alpha'^2)r+2(\alpha'-\beta')\right]\nonumber \\
&&+{\epsilon \kappa \over 1-4\epsilon}(\rho c^2-p_r-2p_t)\,,
\label{eq:Feqs}
\end{eqnarray}
where the prime denotes the derivative with respect to the radial coordinate. Consequently, we obtain the anisotropy parameter, $\Delta(r) = p_t-p_r$, as follows
\begin{equation}\label{eq:Delta1}
\Delta(r) = \frac{e^{-\beta}}{4\kappa r^2}\left[(2\alpha''-\alpha'\beta'+\alpha'^2)r^2-2(\alpha'+\beta')r+4(e^\beta-1)\right].
\end{equation}
Remarkably, the matter-geometry coupling due to the trace $\mathfrak{T}$ does not contribute to the anisotropy once the spherically symmetric spacetime configuration is assumed \citep{Nashed:2022zyi}. However, for $\epsilon \neq 0$ ($\kappa \neq \kappa_E$) slight change in anisotropy is expected. Therefore, deviations from GR due to matter-geometry coupling cannot be spoiled with various anisotropic effects. For the $\epsilon=0$ case, the differential equations \eqref{eq:Feqs} coincide with Einstein field equations of an interior spherically symmetrical spacetime \citep[c.f.,][]{Roupas:2020mvs}.
\subsection{Krori-Barua ansatz}\label{Sec:KB}
We introduce \cite{Krori1975ASS} ansatz (hearafter KB),
\begin{equation}\label{eq:KB}
    \alpha(r)=a_0 x^2+a_1,\,  \beta(r)=a_2 x^2,
\end{equation}
where the dimensionless radius $0 \leq x=r/R \leq 1$ with $R$ being the radius of the star. Additionally, the set of constants \{$a_0, a_1, a_2$\} are dimensionless to be determined by matching conditions on the boundary surface of the star. We further define the dimensionless variables
\begin{equation}
    \bar{\rho}(r)=\frac{\rho(r)}{\rho_{\star}},\, \bar{p}_r(r)=\frac{p_r(r)}{\rho_{\star} c^2},\, \bar{p}_t(r)=\frac{p_t(r)}{\rho_{\star} c^2},\, \bar{\Delta}(r)=\frac{\Delta(r)}{\rho_{\star} c^2},
\end{equation}
where $\rho_{\star}$ denotes a characteristic density
\begin{equation}
    \rho_{\star}=\frac{1}{\kappa_E c^2 R^2}.
\end{equation}
Thus the field equations \eqref{eq:Feqs} read
\begin{eqnarray}
\eta \bar{\rho}&=& \frac{e^{-a_2 x^2}}{x^2}(e^{a_2 x^2}-1+2a_2 x^2) \nonumber\\
&+&\frac{2\epsilon}{x^2}\left[\left(a_0(a_0-a_2)x^4 -(2a_2-3a_0)x^2 +1\right)e^{-a_2x^2} -1\right],\nonumber\\[8pt]
\eta \bar{p}_r&=&\frac{e^{-a_2 x^2}}{x^2}(1-e^{a_2 x^2}+2a_0 x^2) \nonumber\\
&-&\frac{2\epsilon}{x^2}\left[\left(a_0 (a_0-a_2) x^4 - (2 a_2-3 a_0) x^2 +1\right)e^{-a_2 x^2} - 1\right],\nonumber\\[8pt]
\eta \bar{p}_t&=& e^{-a_2 x^2}(2 a_0-a_2 +a_0 (a_0 - a_2) x^2) \nonumber\\
&-&\frac{2\epsilon}{x^2}\left[\left(a_0 (a_0-a_2) x^4 - (2 a_2-3 a_0) x^2 +1\right)e^{-a_2 x^2} - 1\right],\nonumber \\
\label{eq:Feqs2}
\end{eqnarray}
and the anisotropy factor \eqref{eq:Delta1} becomes
\begin{equation}\label{eq:Delta2}
    \eta \bar{\Delta}=\frac{e^{-a_2 x^2}}{x^2}\left[e^{a_2 x^2}-1+a_0(a_0-a_2)x^4 -a_2 x^2\right].
\end{equation}
The mass content within a radius $r$ is given by the mass function
\begin{equation}
    \mathfrak{M}(r)=4 \pi \int_0^r \rho(\bar{r}) \, \bar{r}{^2} dr.
\end{equation}
Substitute the density \eqref{eq:Feqs2} we obtain
\begin{equation}\label{eq:Mass}
    \mathfrak{M}(x)=\frac{M}{C}e^{-a_2 x^2}\left[x(e^{a_2 x^2}-1)+\epsilon \, \zeta(x)\right].
\end{equation}
where the compactness parameter $C$ and the function $\zeta(x)$ are
\begin{eqnarray}
C&=&\frac{2GM}{c^2 R}, \nonumber\\
\zeta(x)&=& a_2^{-\frac{9}{2}}\left\{\left(2 x a_2^\frac{9}{2}-\frac{3}{4} a_0 \sqrt{\pi} a_2^2 (a_2+a_0) \text{erf}(\sqrt{a_2} x) \right) e^{a_2 x^2}\right.\nonumber\\
&+&\left. x \left[\left(\frac{3}{2} a_0 a_2^\frac{5}{2}+a_2^\frac{7}{2} (a_0 x^2 +\frac{3}{2})\right)a_0 -(a_0 x^2+2) a_2^\frac{9}{2}\right]\right\}.\nonumber\\
\end{eqnarray}
We note that the mass function \eqref{eq:Mass} reduces to the GR version where $\epsilon=0$ \citep{Roupas:2020mvs}.
%%%%%%%%%%%%%%%%%%%%%%%%%%%%%%%%%%%%%%%%%%%%%%%%%%%%
\subsection{Matching conditions}\label{Sec:Match}
Since vacuum solutions of both GR and RT are equivalent, the exterior solution is nothing rather Schwarzschild's one,
\begin{equation}
    ds^2=-\left(1-\frac{2GM}{c^2r}\right) c^2 dt^2+\frac{dr^2}{\left(1-\frac{2GM}{c^2 r}\right)}+r^2 (d\theta^2+\sin^2 \theta d\phi^2).
\end{equation}
Recalling the interior spacetime \eqref{eq:metric}, we thus take
\begin{equation}\label{eq:bo}
    \alpha(x=1)=\ln(1-C),\, \beta(x=1)=-\ln(1-C).
\end{equation}
In addition, we make the radial pressure to vanish on the boundary, i.e.
\begin{equation}
    \bar{p}_r(x=1)=0.
\end{equation}
Using KB ansatz \eqref{eq:KB} and the radial pressure \eqref{eq:Feqs2} along with the above the boundary conditions, we derive
\begin{eqnarray}
a_0&=&-\frac{3}{2}-\frac{1}{2}\ln(1-C)-\frac{\sqrt{\chi}+(1-C)}{2\epsilon(1-C)}, \nonumber \\
a_1&=&\frac{3}{2}+\frac{3}{2}\ln(1-C)+\frac{\sqrt{\chi}+(1-C)}{2\epsilon(1-C)}, \nonumber \\
a_2&=&-\ln(1-C),\label{eq:const}
\end{eqnarray}
where
\begin{eqnarray}
    \chi&=&\epsilon^2 (1-C)^2 \ln(1-C)^2 -2\epsilon (1+\epsilon ) (1-C)^2 \ln(1-C) \nonumber\\
    &-&(1-C)[(5C-9)\epsilon^2 -2(2C-3)\epsilon -(1-C)].
\end{eqnarray}
Taking the limit $\epsilon \to 0$, the set of constants \eqref{eq:const} reduces to the GR version \citep{Roupas:2020mvs}
\begin{equation}
    a_0=\frac{C}{2(1-C)},\, a_1=\ln(1-C)-\frac{C}{2(1-C)},\, a_2=-\ln(1-C).
\end{equation}
Interestingly all physical quantities of KB spacetime within a given star, $0\leq x\leq 1$, can be written as dimensionless forms in terms of Rastall and the compactness parameters, i.e. $\bar{\rho}(\epsilon,C)$, $\bar{p}_r(\epsilon,C)$ and $\bar{p}_t(\epsilon,C)$. In practice the compactness parameter is determined by observational data which sets a direct constraint to estimate the nonminimal coupling between matter and geometry as assumed in Rastall gravity. On the other hand it gives a chance to set an upper bound of the allowed compactness for anisotropic neutron star and consequently the maximum mass. This will be discussed in details in Sec. \ref{Sec:MR-reln}.
%%%%%%%%%%%%%%%%%%%%%%%%%%
\subsection{Radial gradients}
We derive the radial gradients of the fluid density, pressures as obtained by Eqs. \eqref{eq:Feqs2}.
\begin{eqnarray}\label{eq:dens_grad}
    \bar{\rho}'&=&\frac{2e^{-a_2 x^2}}{x^3 R}\left\{1- e^{a_2 x^2}-2 a_2^2 x^4+ a_2 x^2-2\epsilon\left[1-e^{a_2 x^2}\right.\right.\nonumber \\
    &-&\left.\left.a_0 a_2(a_2-a_0)x^6-[2(a_2-a_0)^2-a_0^2] x^4+a_2 x^2\right]\right\},\nonumber\\
\end{eqnarray}
\begin{eqnarray}\label{eq:pr_grad}
    \bar{p}'_r&=&\frac{-2e^{-a_2 x^2}}{x^3 R}\left\{1- e^{a_2 x^2}+2 a_0 a_2 x^4+ a_2 x^2-2\epsilon\left[1-e^{a_2 x^2}\right.\right.\nonumber \\
    &-&\left.\left.a_0 a_2(a_2-a_0)x^6-[2(a_2-a_0)^2-a_0^2] x^4+a_2 x^2\right]\right\},
\end{eqnarray}
\begin{eqnarray}\label{eq:pt_grad}
    &&\bar{p}'_t=\frac{2e^{-a_2 x^2}}{x^3 R}\left\{a_0 a_2(a_2-a_0) x^6+ [a_2(a_2-3 a_0)+a_0^2] x^4+2\epsilon\, \times\right.\nonumber \\
    &&\left.\left[1-e^{a_2 x^2}-a_0 a_2(a_2-a_0)x^6-[2(a_2-a_0)^2-a_0^2] x^4+a_2 x^2\right]\right\},\nonumber\\
\end{eqnarray}
where $'=R^{-1} d/dx$. Those gradients are of a great importance since they determine the physical stability of the compact object as will be discussed in the following section.
%%%%%%%%%%%%%%%%%%%%%%%%%%%%%%%%%%%%%%%%%%%%%% Section 4 %%%%%%%%%%%%%%%%%%%%%%%%%%%%%%%%%%%%%%%%%%%%%%%%%%%%%%%%%%%%%%%
\section{Observational Constraints and Stability Conditions}\label{Sec:Stability}
In this section we use the precise X-ray mass and radius constraints on the pulsar PSR J0740+6620 from NICER+XMM based on Gaussian process applying a nonparameteric EoS approach, the mass and radius are given as $M=2.07 \pm 0.11 M_\odot$ and $R=12.34^{+1.89}_{-1.67}$ km \cite{Legred:2021hdx} where the solar mass $M_\odot=1.9891\times 10^{30}$ kg. For a stellar model to be physically well behaved, it needs to satisfy the conditions as labeled from (\textbf{i}) to (\textbf{xii}) as follows:
%%%%%%%%%%%%%%%%%%%%%%%%%%%
\subsection{Zeldovich condition and Rastall parameter}
\noindent Condition (\textbf{i}): According to \cite{1971reas.book.....Z} the central radial pressure must be less than or equal to the central density, then we write
\begin{equation}\label{eq:Zel}
    \frac{\bar{p}_r(0)}{\bar{\rho}(0)}\leq 1.
\end{equation}
We obtain the density, the radial and the tangential pressure at the center
\begin{eqnarray}
% \nonumber to remove numbering (before each equation)
  \bar{\rho}(x=0) &=& 6(a_0-a_2)\epsilon+3 a_2 \nonumber \\
  \bar{p}_r(x=0)   &=& -6(a_0-a_2)\epsilon-a_2+2 a_0= \bar{p}_t(x=0).\quad
\end{eqnarray}
For the pulsar PSR J0740+6620 we evaluate its compactness parameter $C = 0.510\pm 0.0999$. Then, by applying Zeldovich condition \eqref{eq:Zel} we choose a reasonable valid interval of Rastall parameter $0\leq \epsilon \leq 0.278\pm 0.0067$ where $\epsilon$ is expected to be close to the GR version ($\epsilon=0$).
%%%%%%%%%%%%%%%%%%%%%%%%%%%%%%%%%%%%%%%%%%
\subsection{Mass and radius observational constraints from PSR J0740+6620}
\begin{figure}
\centering
{\includegraphics[scale=0.3]{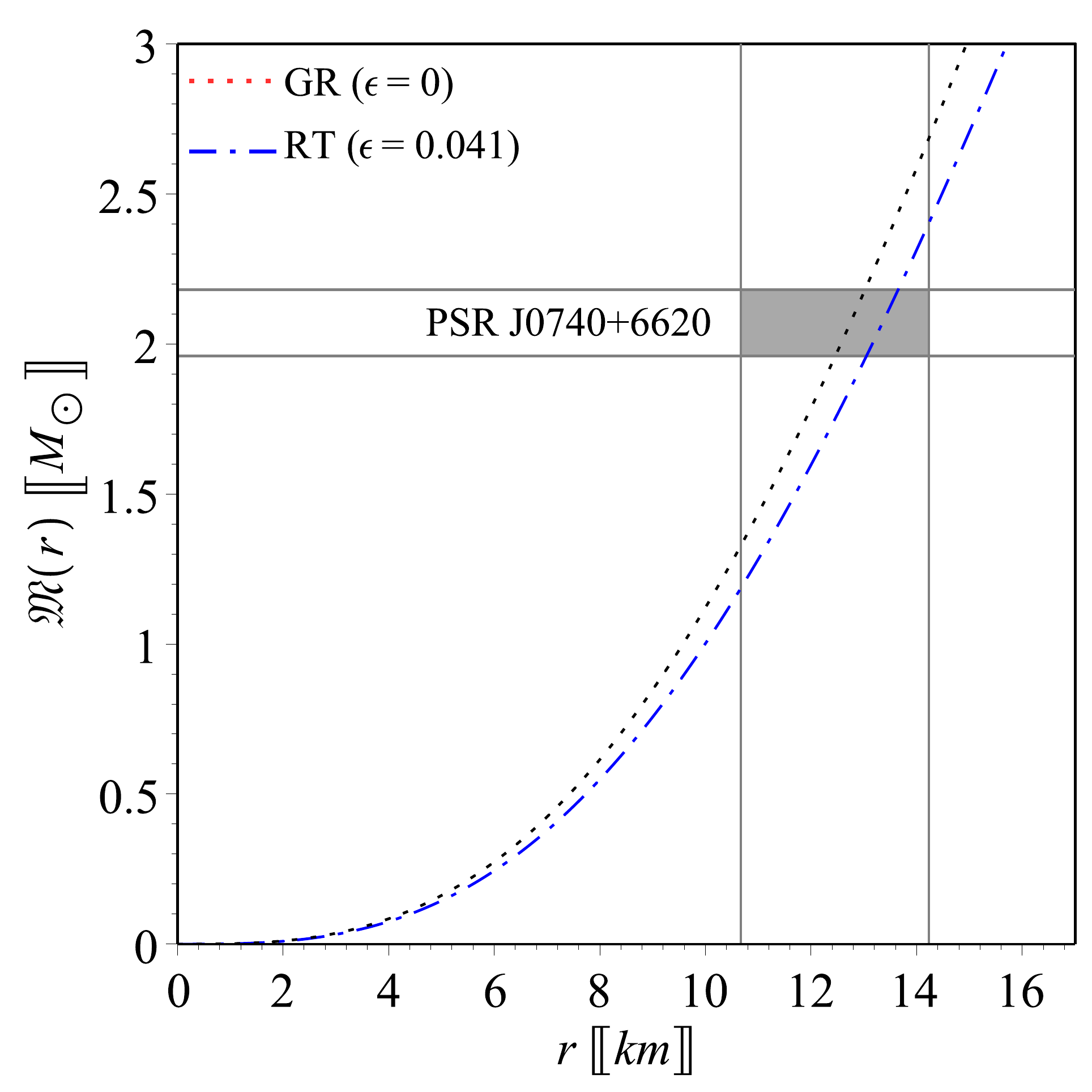}}
\caption{The plot of the mass function \eqref{eq:Mass} of the pulsar PSR J0740+6620. The shaded area represents the feasible region as imposed by NICER+XMM observational data ($M=2.07 \pm 0.11 M_\odot$ and $R=12.34^{+1.89}_{-1.67}$ km) \cite{Legred:2021hdx}. For Rastall case we use \{$\epsilon=0.041$, $\kappa=2.302\times 10^{-43}\,N^{-1}$, $C=0.491$, $a_0 = 0.438$, $a_1 = -1.114$, $a_2 =0.676$\}. For GR case ($\epsilon=0$) the corresponding values are \{$\kappa = \kappa_E= 2.077\times 10^{-43}\,N^{-1}$, $a_0 = 0.491$, $a_1 = -1.176$, $a_2 = 0.684$\}.}
\label{Fig:Mass}
\end{figure}
Recalling the X-ray mass and radius constraints on the pulsar PSR J0740+6620 from NICER+XMM, the mass function \eqref{eq:Mass} estimates a total mass $M=1.96 M_\odot$ at a radius of $R = 13.04$ km with compactness $C=0.491$ in agreement with the observed value, ($M=2.07 \pm 0.11 M_\odot$ and $R=12.34^{+1.89}_{-1.67}$ km) \cite{Legred:2021hdx}, for a choice of Rastall parameter $\epsilon=0.041$. This determines the coupling constant, \eqref{Rconst}, and the set of constants, Eqs. \eqref{eq:const}, \{$\kappa=2.302\times 10^{-43}\,N^{-1}$, $a_0 = 0.438$, $a_1 = -1.114$, $a_2 =0.676$\}. These values clearly satisfy Zeldovich condition \eqref{eq:Zel}. We plot the mass function pattern in Fig. \ref{Fig:Mass} to show the agreement of the predicted mass-radius of the pulsar PSR J0740+6620 and their observed values as obtained by NICER+XMM data. In comparison with the GR prediction, RT with positive $\epsilon$ estimates the same mass as in GR but within larger size star (or less mass at same size). On another word, RT predicts compactness value lesser than GR for a given mass. This reflects the capability of RT to allow for more masses or equivalently higher values of compactness while the stability conditions are being still fulfilled. This will be discussed in more details in Sec. \ref{Sec:MR-reln}. Notably, for $\epsilon < 0$, the matter-geometry coupling turns the star to slightly smaller sizes in comparison with GR for the same mass \citep{Nashed:2022zyi}; and therefore we omit this choice in the present work.
%%%%%%%%%%%%%%%%%%%%%%%%%%%
\subsection{Geometric sector}\label{Sec:geom}
\noindent Condition (\textbf{ii}): For the geometric sector, the metric potentials $g_{tt}$ and $g_{rr}$ should be free from coordinate and physical singularities within the interior region of the star $0\leq r\leq R$, where the center (boundary) is at $r=0$ ($r=R$) respectively. Obviously the metric \eqref{eq:metric} satisfies these conditions whereas at the center, $g_{tt}(x=0)=-e^{a_1}$ and $g_{rr}(x=0)=1$, and they are finite everywhere inside the star $0 \leq x \leq 1$.

\noindent Condition (\textbf{iii}): The metric potentials of the interior solution and the exterior should match smoothly at the boundary.
\begin{figure}
\centering
{\includegraphics[scale=0.3]{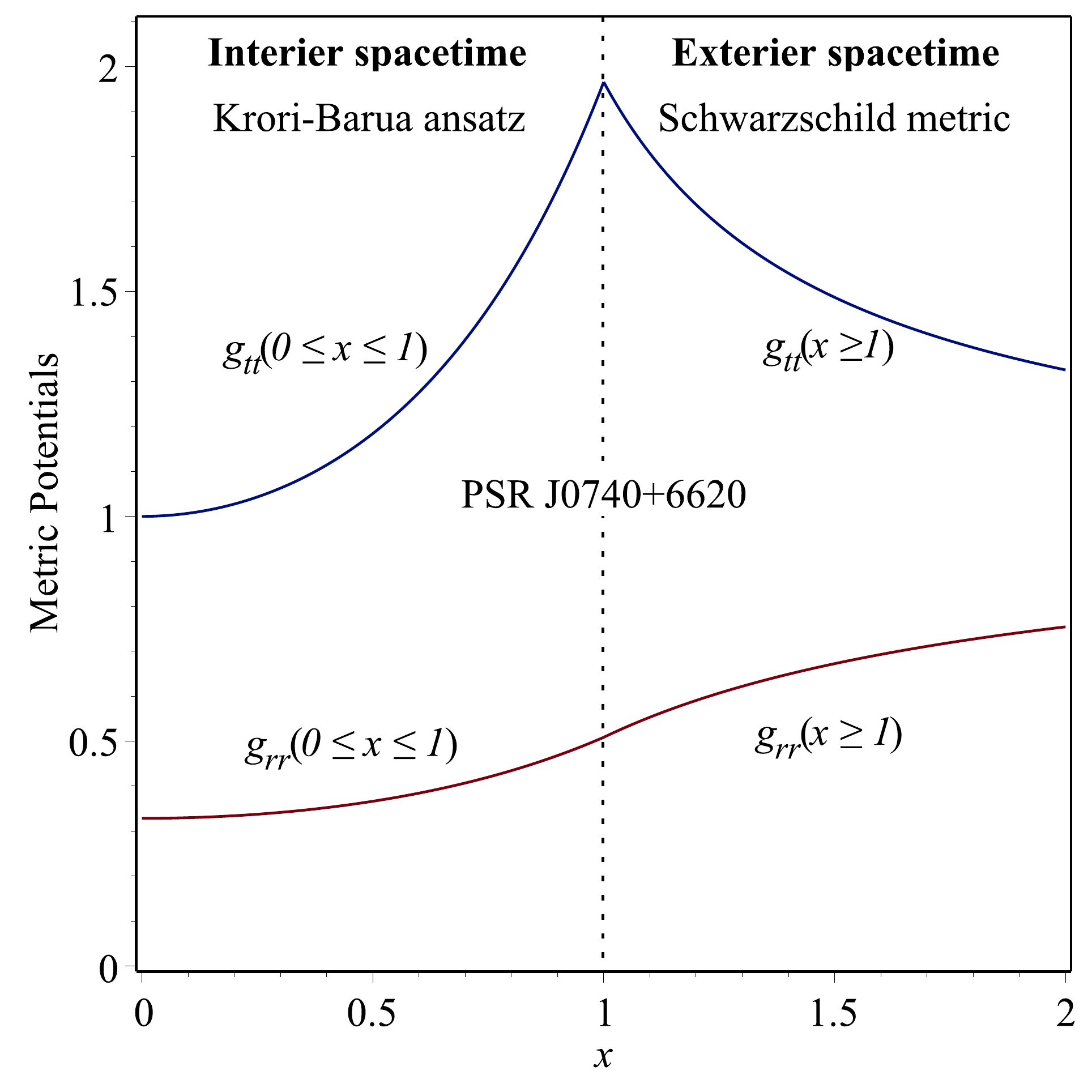}}
\caption{The KB potentials of the pulsar PSR J0740+6620 are finite everywhere inside the star and match smoothly Schwarzschild exterior vacuum solution. The plot confirms that conditions (\textbf{ii}) and (\textbf{iii}) are satisfied.}
\label{Fig:Matching}
\end{figure}

\textit{Obviously conditions} (\textbf{ii}) and (\textbf{iii}) \textit{are satisfied for the pulsar PSR J0740+6620 as obtained by Fig. \ref{Fig:Matching}}.

\noindent Condition (\textbf{iv}): We define the gravitational red-shift of the metric \eqref{eq:metric}
\begin{equation}\label{eq:redshift}
    Z=\frac{1}{\sqrt{-g_{tt}}}-1=\frac{1}{\sqrt{e^{a_0 x^2+a_1}}}-1.
\end{equation}
The redshift should be finite and positive everywhere inside the star and decreases monotonically toward the boundary, i.e. $Z>0$ and $Z'<0$.
\begin{figure}
\centering
{\includegraphics[scale=0.3]{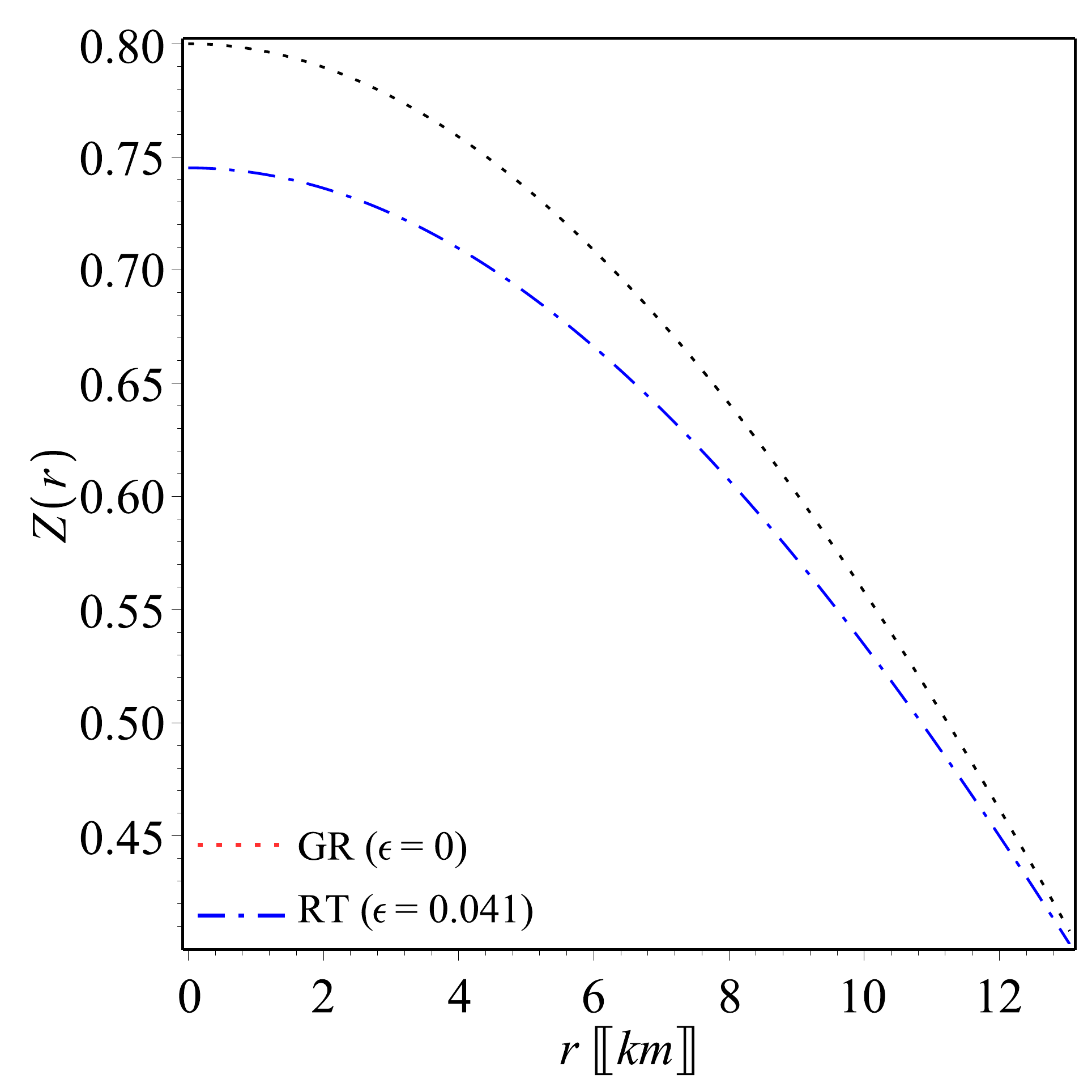}}
\caption{The redshift function \eqref{eq:redshift} of the pulsar PSR J0740+6620. The plot confirms that condition (\textbf{iv}) is satisfied.}
\label{Fig:Redshift}
\end{figure}

\textit{Obviously condition} (\textbf{iv}) \textit{is satisfied for the pulsar PSR J0740+6620 as obtained by Fig. \ref{Fig:Redshift}}.

At the center $Z(0)\approx 0.75$ (less than the GR value $Z(0)\approx 0.8$) is maximum while at the surface $Z_R\approx 0.4$ (similar to the GR value) less than the upper redshift bound $Z_R=2$ as obtained by \cite{PhysRev.116.1027}, see also \citep{Ivanov:2002xf,PhysRevD.67.064003} for anisotropic case and \cite{Bohmer2006} in presence of a cosmological constant. We note that the maximal redshift constraint is not a very strict condition, subsequently it cannot provide a proper way to put an upper bound on the compactness as already concluded by some work \citep[c.f.,][]{Ivanov:2002xf,PhysRevD.67.064003,Bohmer2006}. This conclusion is altered when using the energy conditions on the matter sector which provide much better constraints, c.f. \cite{Roupas:2020mvs}.
%%%%%%%%%%%%%%%%%%%%%%%%%%%%
\subsection{Matter sector}\label{Sec:matt}
\begin{figure*}
\centering
\subfigure[~The energy-density]{\label{fig:density}\includegraphics[scale=0.25]{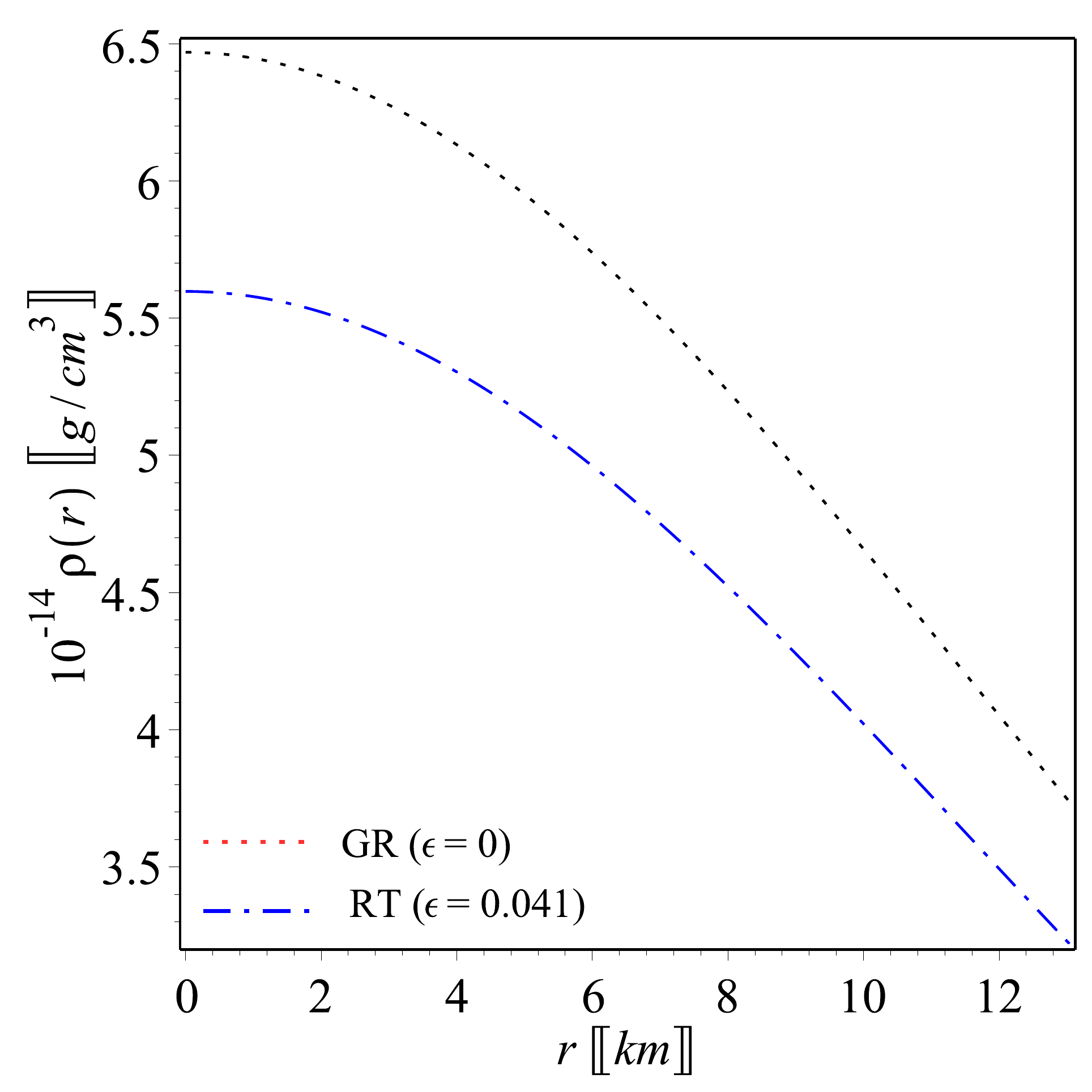}}
\subfigure[~The radial pressure]{\label{fig:radpressure}\includegraphics[scale=0.25]{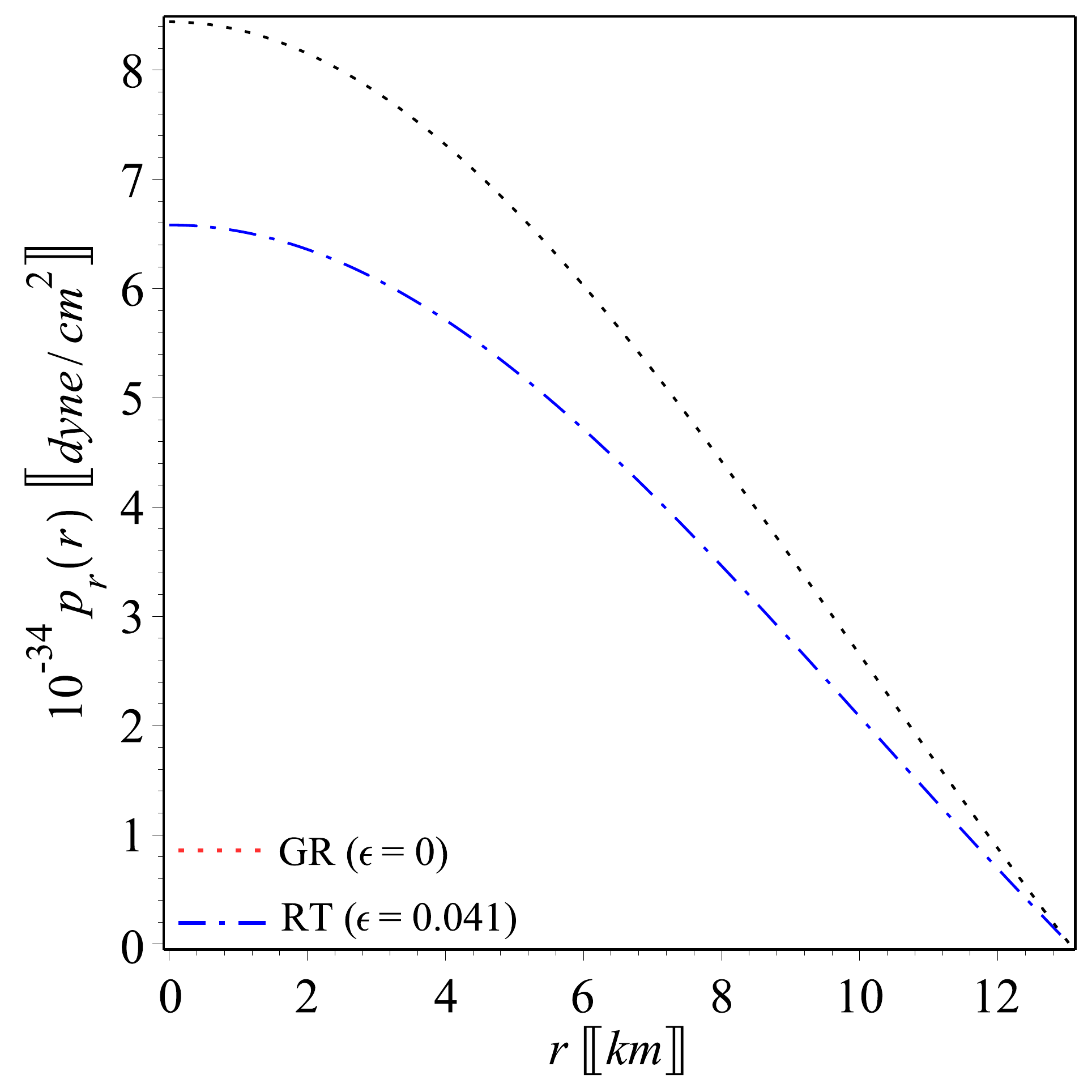}}
\subfigure[~The tangential pressure]{\label{fig:tangpressure}\includegraphics[scale=0.25]{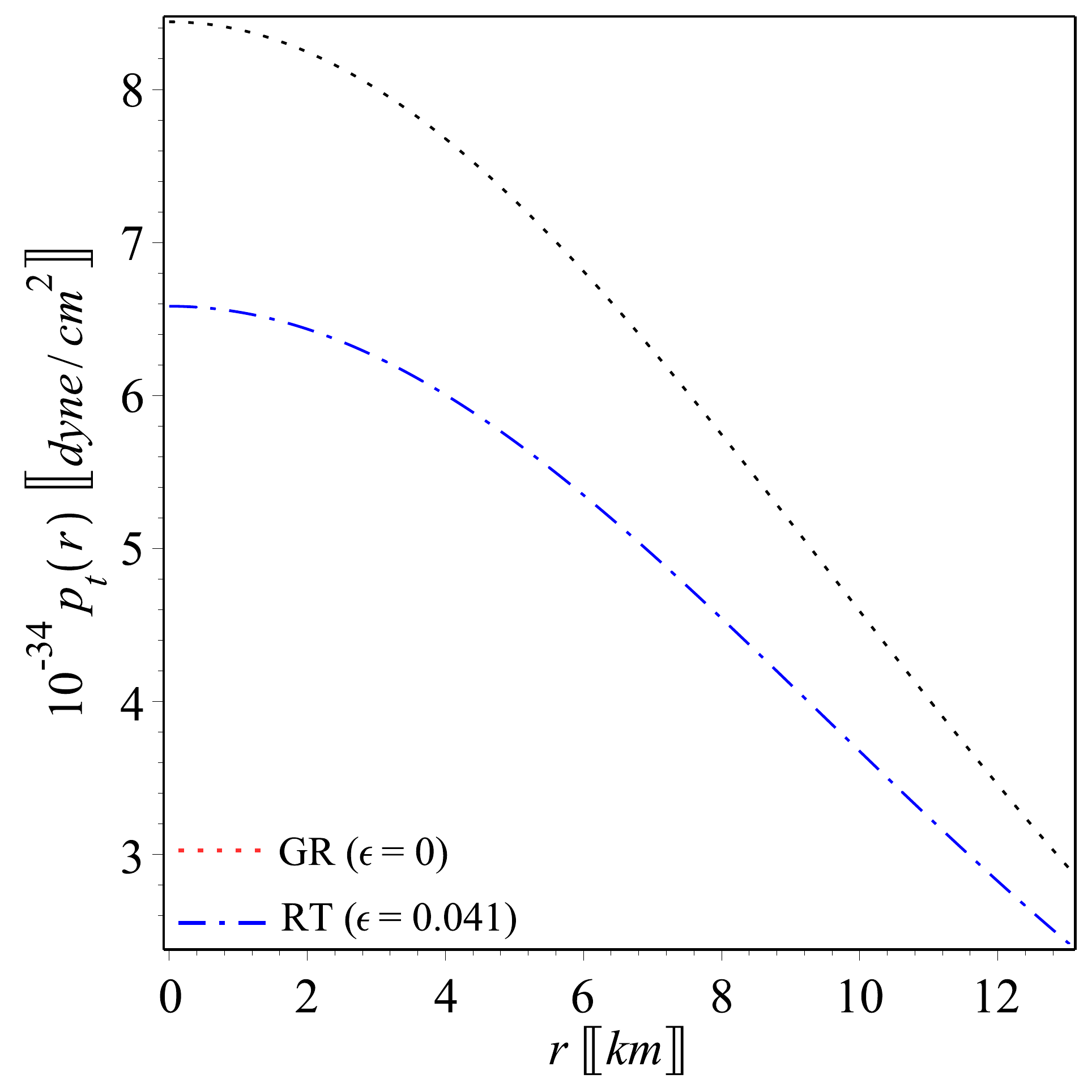}}\\
\subfigure[~The gradients (GR)]{\label{fig:GRgrad}\includegraphics[scale=0.25]{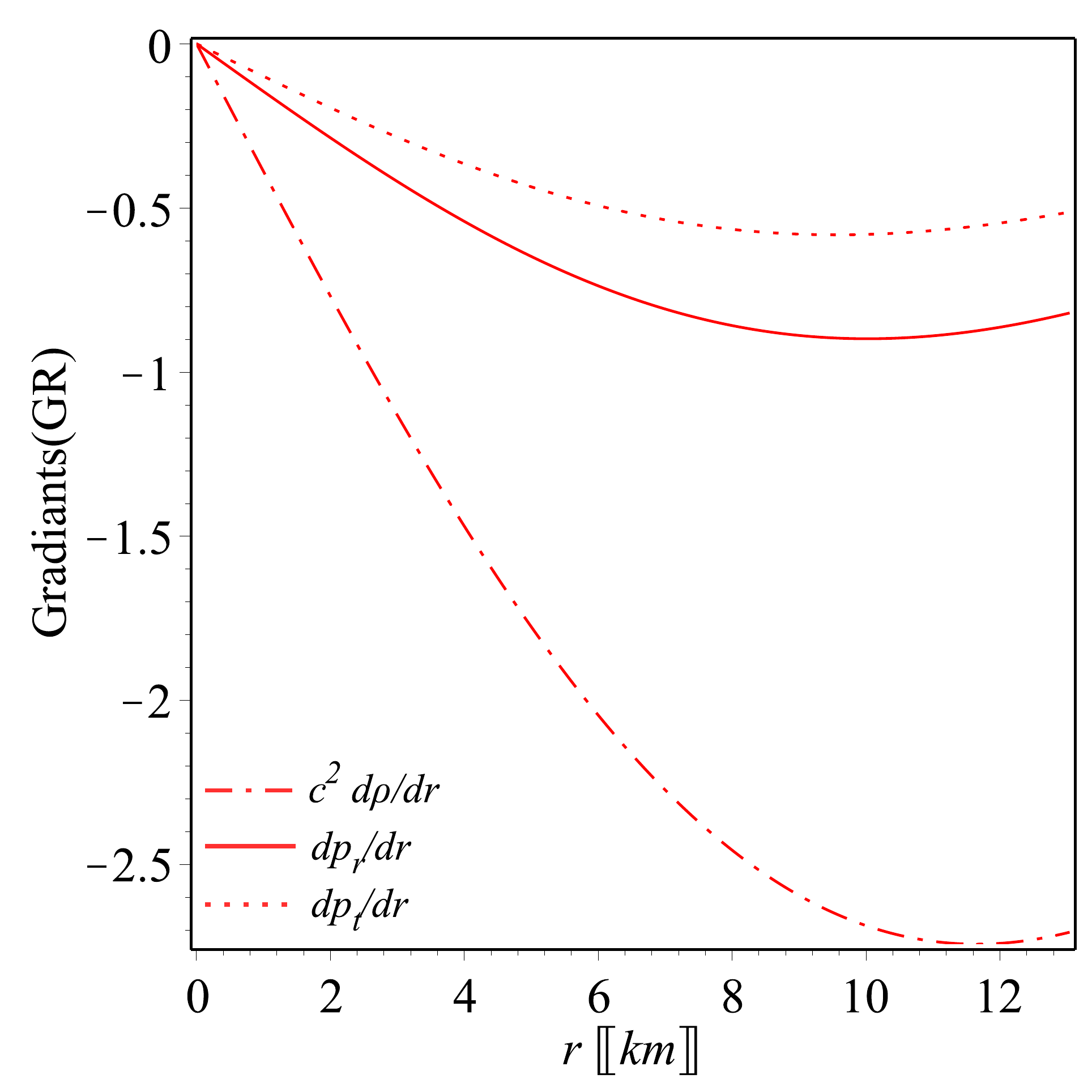}}
\subfigure[~The gradients (RT)]{\label{fig:RTgrad}\includegraphics[scale=0.25]{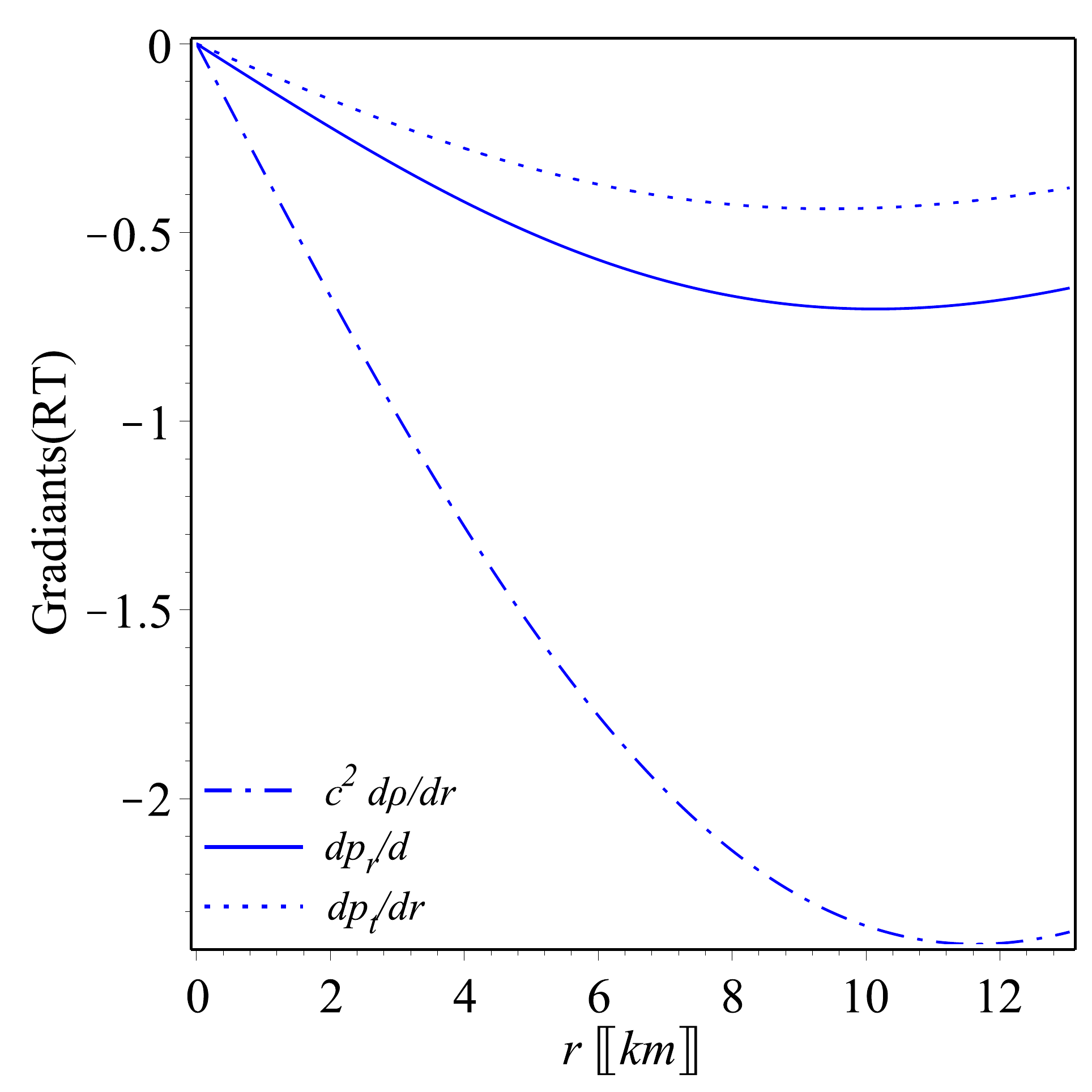}}
\subfigure[~The anisotropy parameter]{\label{fig:anisot}\includegraphics[scale=0.25]{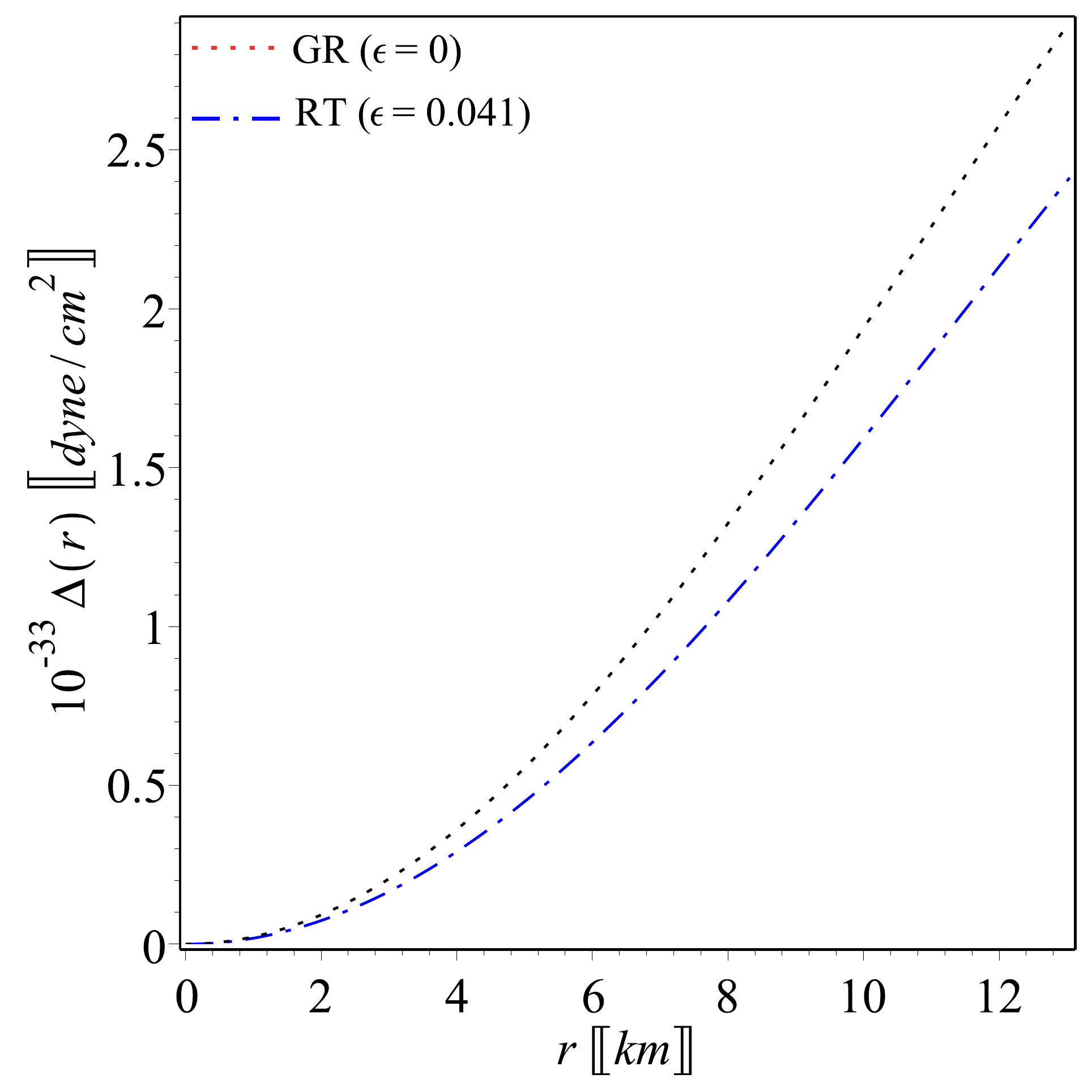}}
\caption{The patterns of the density, radial and tangential pressures, Eqs. \eqref{eq:Feqs2}, of the pulsar PSR J0740+6620 are given by Figs. \subref{fig:density}--\subref{fig:tangpressure}. The radial gradients of these quantities are given by Figs. \subref{fig:GRgrad} and \subref{fig:RTgrad} for GR and RT cases. \subref{fig:anisot} the plots of the anisotropy parameter $\Delta$, Eq. \eqref{eq:Delta2}. The plots confirm that conditions (\textbf{v})--(\textbf{vii}) are satisfied.}
\label{Fig:dens_press}
\end{figure*}
\noindent Condition (\textbf{v}):  For a solution to be regular, the density, radial and tangential pressures of the fluid should be non singular everywhere inside the star. Additionally, these physical quantities should have maximum values at the center and monotonically decrease towards the surface of the star. i.e.
\begin{itemize}
  \item[a.] $\rho(r=0)>0$, $\rho'(r=0)=0$, $\rho''(r=0)<0$ and $\rho'(0< r \leq R)< 0$,
  \item[b.] $p_r(r=0)>0$, $p_r'(r=0)=0$, $p_r''(r=0)<0$ and $p_r'(0< r \leq R)< 0$,
  \item[c.] $p_t(r=0)>0$, $p_t'(r=0)=0$, $p_t''(r=0)<0$ and $p_t'(0< r \leq R)< 0$.
\end{itemize}

\noindent Condition (\textbf{vi}): The density, radial and tangential pressures of the fluid within the star ($0 < r < R$), should be positive, i.e. $\rho(0 < r < R)>0$, $p_r(0 < r < R)>0$ and $p_t(0 < r < R)>0$.

\noindent Condition (\textbf{vii}): The radial pressure of the fluid should vanish at the boundary surface of the star, i.e. $p_r(r=R)=0$. This is not necessarily for the tangential pressure.

\textit{Obviously conditions} (\textbf{v})--(\textbf{vii}) \textit{are satisfied for the pulsar PSR J0740+6620 as obtained by Fig. \ref{Fig:dens_press}\subref{fig:density}--\subref{fig:RTgrad}}.

We note that the model estimates a NS core density $\rho_\text{core}\approx 5.6\times 10^{14}$ g/cm$^{3} \approx 2.1\rho_\text{nuc}$ of the pulsar PSR J0740+6620, which does not exclude the possibility of the pulsar core being neutrons. In addition, at this density the anaisotropic fluid would be a realistic assumption.

\noindent Condition (\textbf{vii}): The anisotropy parameter $\Delta$ should vanish at the center of the star, i.e., $p_r(r=0)=p_t(r=0)$, increasing toward the boundary, i.e. $\Delta'(0 \leq r\leq R)>0$. Consequently the anisotropic force $F_a=2\Delta/r$ vanishes at the center. Taking the limit $x\to 0$ in Eq. \eqref{eq:Delta2} we obtain $\Delta\to 0$.

\textit{Obviously condition} (\textbf{vii}) \textit{is satisfied for the pulsar PSR J0740+6620 as obtained by Fig. \ref{Fig:dens_press}\subref{fig:anisot}}.

Also, one finds that the anaistropy parameter $\Delta(r>0) > 0$ (i.e., $p_t>p_r$) which is necessarily for the induced anaisotropic force to be repulsive, and therefore allows for larger NS size in comparison with the isotropic perfect fluid case. However, for $\epsilon>0$, the anisotropy in RT is slightly less than GR due to larger coupling constant $\kappa$ of Rastall relative to Einstein one $\kappa_E$ as obtained by Eq. \eqref{eq:Delta1}.
%%%%%%%%%%%%%%%%%%%%%%%%%%%%%%%%%%%%%%
\subsection{Energy conditions}\label{Sec:Energy-conditions}
In the context of GR, the focusing theorem implies the positivity of the tidal tensor trace $\mathfrak{R}_{\alpha\beta} u^{\alpha} u^{\beta} \geq 0$ and $\mathfrak{R}_{\alpha\beta} \ell^{\alpha} \ell^{\beta} \geq 0$ in Raychaudhuri equation, where $u^{\alpha}$ is an arbitrary timelike vector and $\ell^{\alpha}$ is an arbitrary future directed null vector. This imposes four conditions on the energy-momentum tensor $\mathfrak{T}^{\alpha\beta}$, those are the energy conditions. These could be extended to modified gravity. In the particular case of Rastall's gravity the energy conditions could be written in terms of the effective energy-momentum tensor $\widetilde{\mathfrak{T}}{^\alpha}{_\beta}=diag(-\tilde{\rho}c^2,\tilde{p}_r, \tilde{p}_t, \tilde{p}_t)$, since $\mathfrak{R}_{\alpha\beta}=\kappa\left(\widetilde{\mathfrak{T}}_{\alpha\beta}-\frac{1}{2} g_{\alpha\beta} \widetilde{\mathfrak{T}}\right)$.
\begin{figure*}[t!]
\centering
\subfigure[Dominance of density over radial pressure]{\label{fig:Cond1}\includegraphics[scale=0.29]{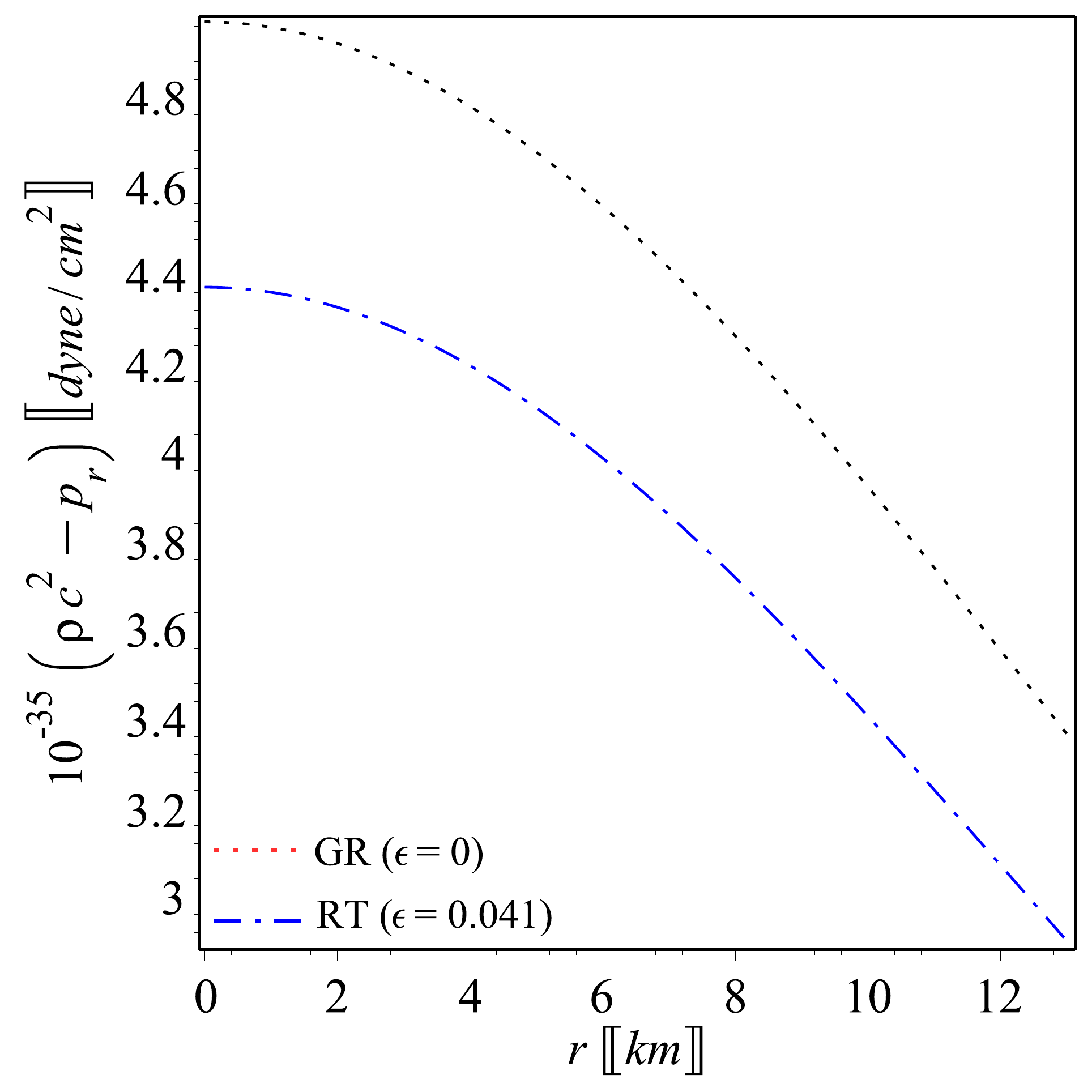}}\hspace{0.1cm}
\subfigure[Dominance of density over tangential pressure]{\label{fig:Cond2}\includegraphics[scale=0.29]{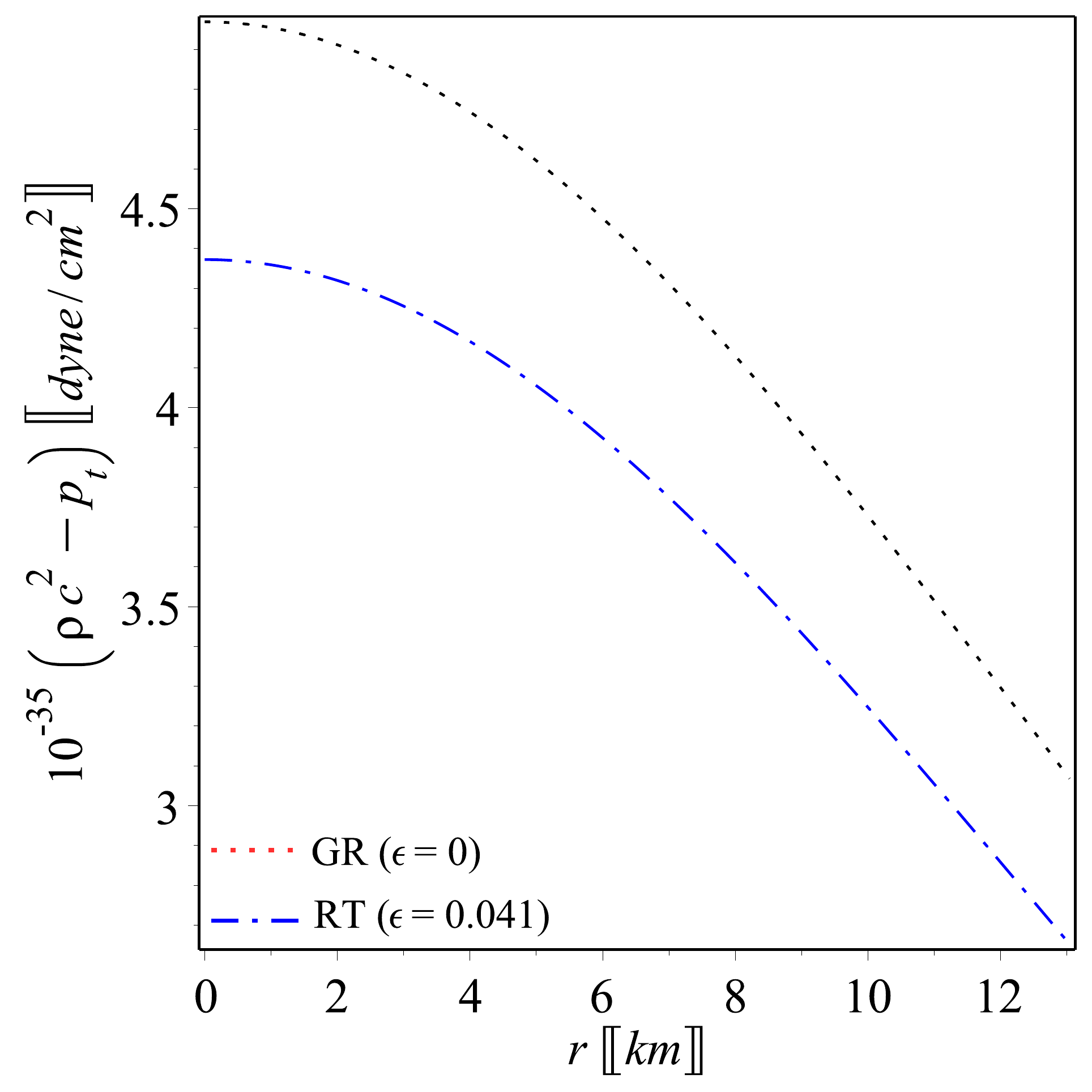}}\hspace{0.1cm}
\subfigure[Dominance of density over total pressure]{\label{fig:Cond3}\includegraphics[scale=.29]{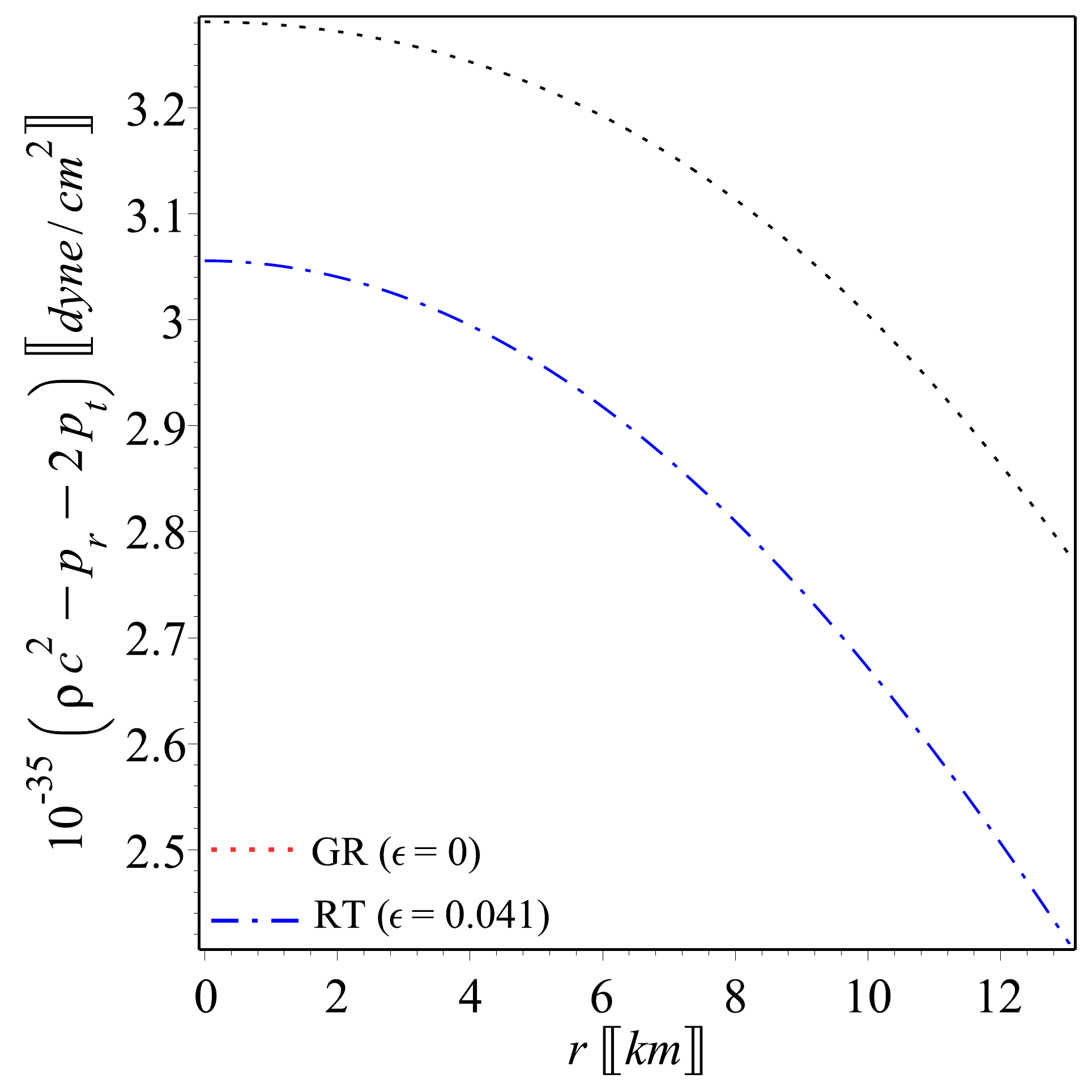}}
%\subfigure[~Dominant energy condition (radial)]{\label{fig:DEC}\includegraphics[scale=.23]{Fig5d.pdf}}\hspace{2cm}
%\subfigure[~Dominant energy condition (tangential)]{\label{fig:DEC2}\includegraphics[scale=.23]{Fig5e.pdf}}
%
\caption{The patterns of the energy conditions according to the solution \eqref{eq:Feqs} of the pulsar PSR J0740+6620. The plots confirm that energy conditions (\textbf{viii}) all are satisfied.}
\label{Fig:EC}
\end{figure*}

\noindent Condition (\textbf{viii}): A physical model should satisfy the modified energy conditions as stated below:
\begin{itemize}
  \item[a.] Weak energy condition (WEC): $\tilde{\rho}\geq 0$, $ \tilde{\rho} c^2+ \tilde{p}_r > 0$, $\tilde{\rho} c^2+\tilde{p}_t > 0$,
  \item[b.] Null energy condition (NEC): $\tilde{\rho} c^2+ \tilde{p}_r \geq 0$, $\tilde{\rho} c^2+  \tilde{p}_t \geq 0$,
  \item[c.] Strong energy condition (SEC): $\tilde{\rho} c^2+\tilde{p}_r+2\tilde{p}_t \geq 0$, $\tilde{\rho} c^2+\tilde{p}_r \geq 0$, $\tilde{\rho} c^2+\tilde{p}_t \geq 0$,
  \item[d.] Dominant energy conditions (DEC): $\tilde{\rho}\geq 0$, $\tilde{\rho} c^2-\tilde{p}_r \geq 0$ and $\tilde{\rho} c^2-\tilde{p}_t \geq 0$.\\
\end{itemize}
We rewrite the field equations \eqref{eq:Feqs} in the following form
\begin{eqnarray}\label{eq:Eff-dens-press}
\nonumber \tilde{\rho} c^2 &=& \rho c^2 + \frac{\epsilon}{1-4\epsilon} (\rho c^2 - p_r - 2p_t),\\
\nonumber \tilde{p}_r &=& p_r - \frac{\epsilon}{1-4\epsilon} (\rho c^2 - p_r - 2p_t),\\
\tilde{p}_t &=& p_t - \frac{\epsilon}{1-4\epsilon} (\rho c^2 - p_r - 2p_t).
\end{eqnarray}
For $\epsilon=0$ the GR is restored. It is straightforward to prove that once the NEC is fulfilled in GR, it is also fulfilled in RT whereas $\tilde{\rho} c^2 + \tilde{p}_r=\rho c^2 + p_r$ and $\tilde{\rho} c^2 + \tilde{p}_t=\rho c^2 + p_t$. By virtue of conditions  (\textbf{v}) and  (\textbf{vi}), explicitly the density and the pressures are positive within the star, the NEC is satisfied. Additionally, we show that
\begin{eqnarray}\label{eq:Ras_EC}
\nonumber \tilde{\rho} c^2- \tilde{p}_r &=& (\rho c^2 - p_r) + \frac{2\epsilon}{1-4\epsilon} (\rho c^2 - p_r - 2p_t),\\
\nonumber \tilde{\rho} c^2- \tilde{p}_t &=& (\rho c^2 - p_t) + \frac{2\epsilon}{1-4\epsilon} (\rho c^2 - p_r - 2p_t),\\
\tilde{\rho} c^2 + \tilde{p}_r + 2\tilde{p}_t &=& (\rho c^2 + p_r + 2p_t) - \frac{2\epsilon}{1-4\epsilon} (\rho c^2 - p_r - 2p_t).\qquad
\end{eqnarray}
Given that the fraction $\frac{\epsilon}{1-4\epsilon}>0$ for $0 < \epsilon <1/4$ (in our case $\epsilon=0.041$) in addition to $\rho\geq 0$, $p_r\geq 0$ and $p_t\geq 0$; it remains to show that $\rho c^2 - p_r\geq 0$, $\rho c^2- p_t\geq 0$ and $\rho c^2 - p_r - 2p_t \geq 0$ to grantee the verification of the energy conditions (a--d).

\textit{Obviously the energy conditions} (\textbf{viii}) \textit{are satisfied for the pulsar PSR J0740+6620 as obtained by Fig. \ref{Fig:EC}\subref{fig:Cond1}--\subref{fig:Cond3}}.

Finally we note that the dominance of energy density over the total pressure is guaranteed in RT as long as it is fulfilled in GR, since $\tilde{\rho} c^2 - \tilde{p}_r - 2\tilde{p}_t = \frac{1}{1-4\epsilon}(\rho c^2 - p_r - 2p_t)$ whereas $0 \leq \epsilon < 1/4$. This latter constraint has been referred to as a strong energy condition by some authors \citep[c.f.][]{1988CQGra...5.1329k,Ivanov:2017kyr,2019EPJC...79..853D,Roupas:2020mvs}. However, we will call it as a dominant energy condition keeping in mind that imposed constraints by this condition for matter or effective fluids are identical.

According to the above results one can see that the energy conditions for the effective fluid are strongly related to the matter fluid in RT, but not identical in all cases as expected, \citep[c.f.][]{Moradpour:2016ubd,2019MNRAS.486.2407L}. However, the DEC in the matter sector $\rho c^2 - p_r - 2p_t\geq 0$ is the core condition which allows for other conditions to be satisfied in general whereas the matter density and pressures are positive and $0 < \epsilon < 1/4$. Similar constraints on Rastall parameter motivated by thermodynamics aspects (positivity of the horizon entropy) have been obtained by \cite{Moradpour:2016fur}. Furthermore, in cosmological applications, it has been shown that the second law of thermodynamics is fulfilled in RT, if the WEC is verified for the matter sector \citep{2016PhLB..757..187M}.
%%%%%%%%%%%%%%%%%%%%%%%%%%%%%%%%%%%%%%%%%%%%%%%%%%%%%
\subsection{Causality and stability conditions}
\begin{figure*}
\centering
\subfigure[~Radial speed of sound]{\label{fig:vr}\includegraphics[scale=0.28]{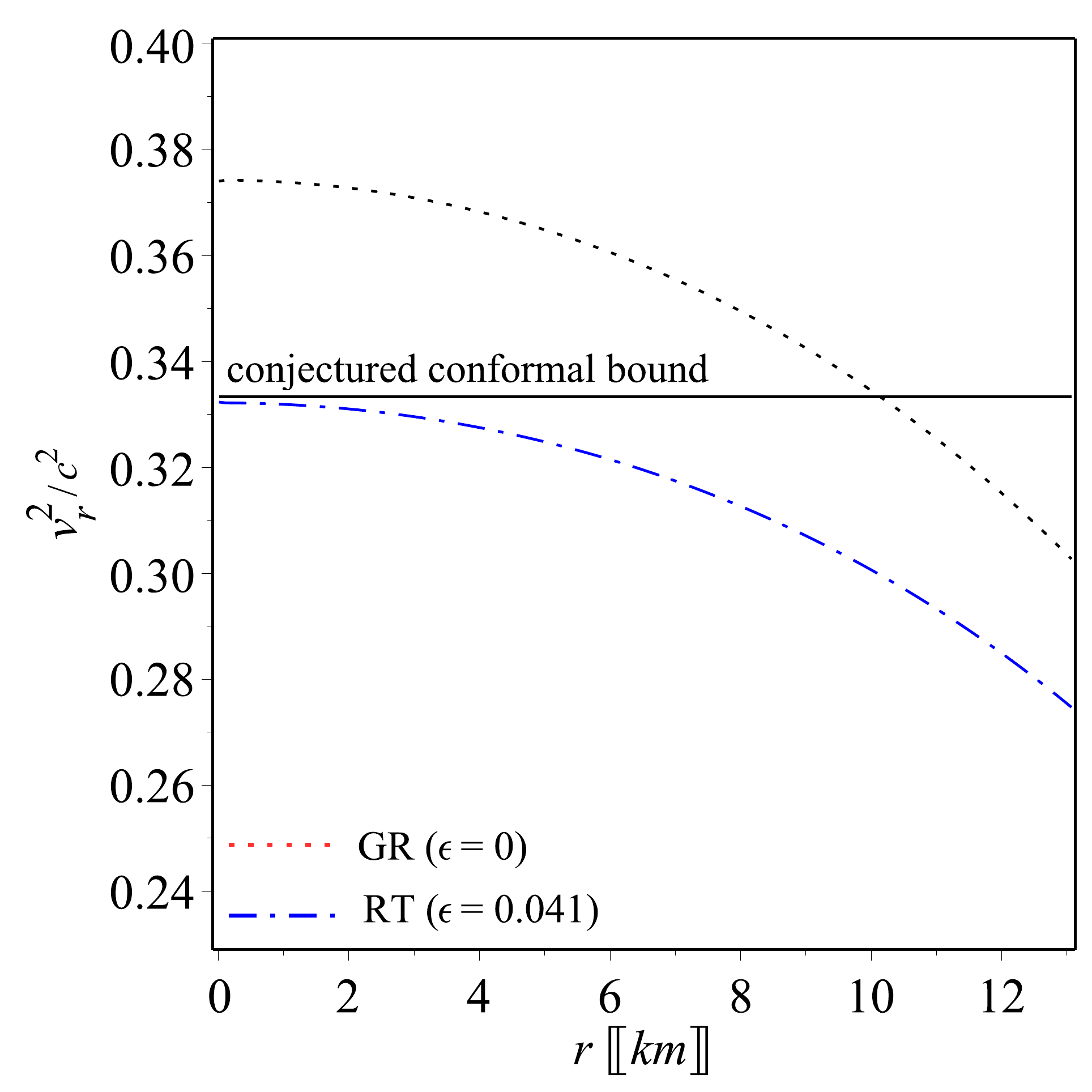}}
\subfigure[~Tangential speed of sound]{\label{fig:vt}\includegraphics[scale=.28]{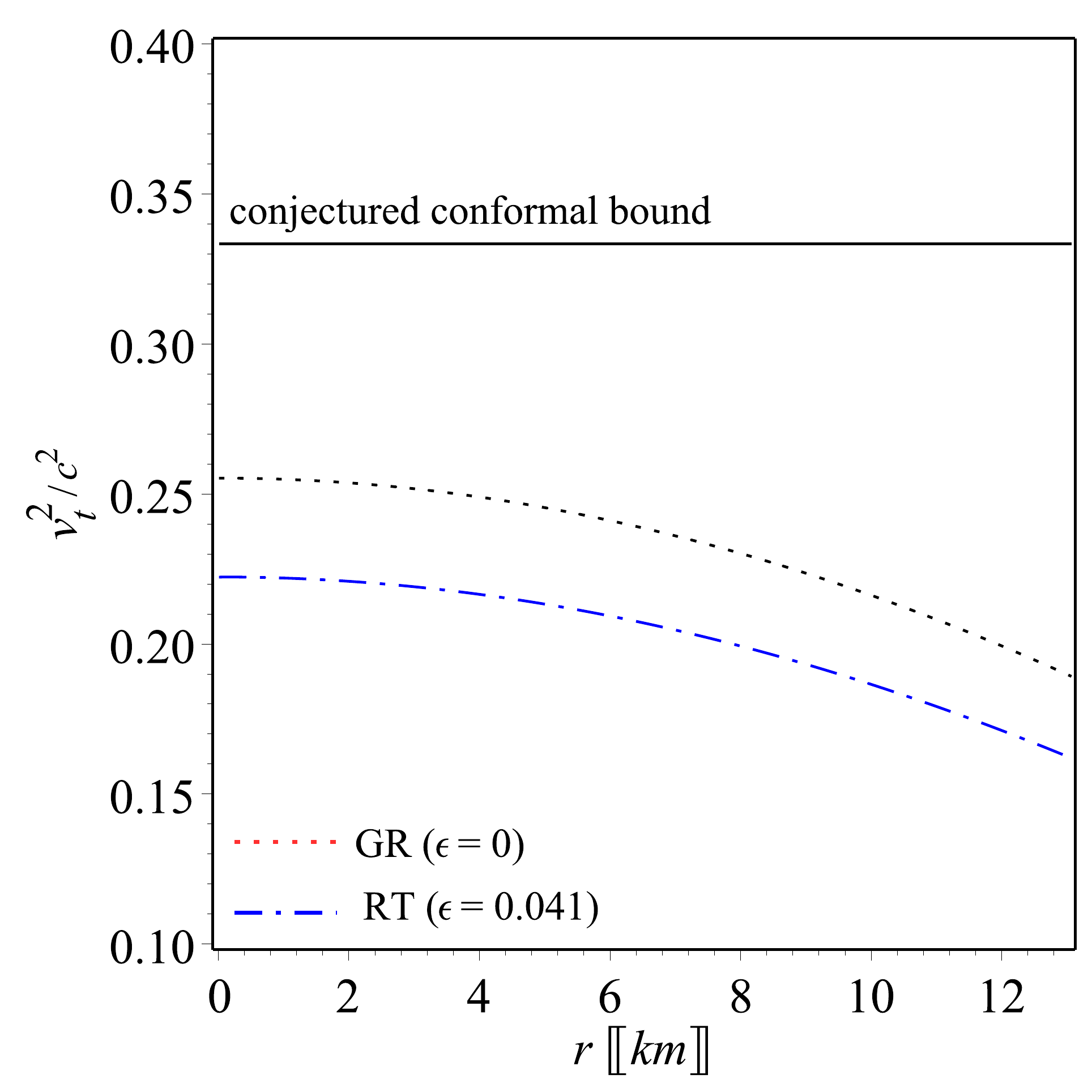}}
\subfigure[~difference between radial and tangential  speed of sounds]{\label{fig:vt-vr}\includegraphics[scale=.28]{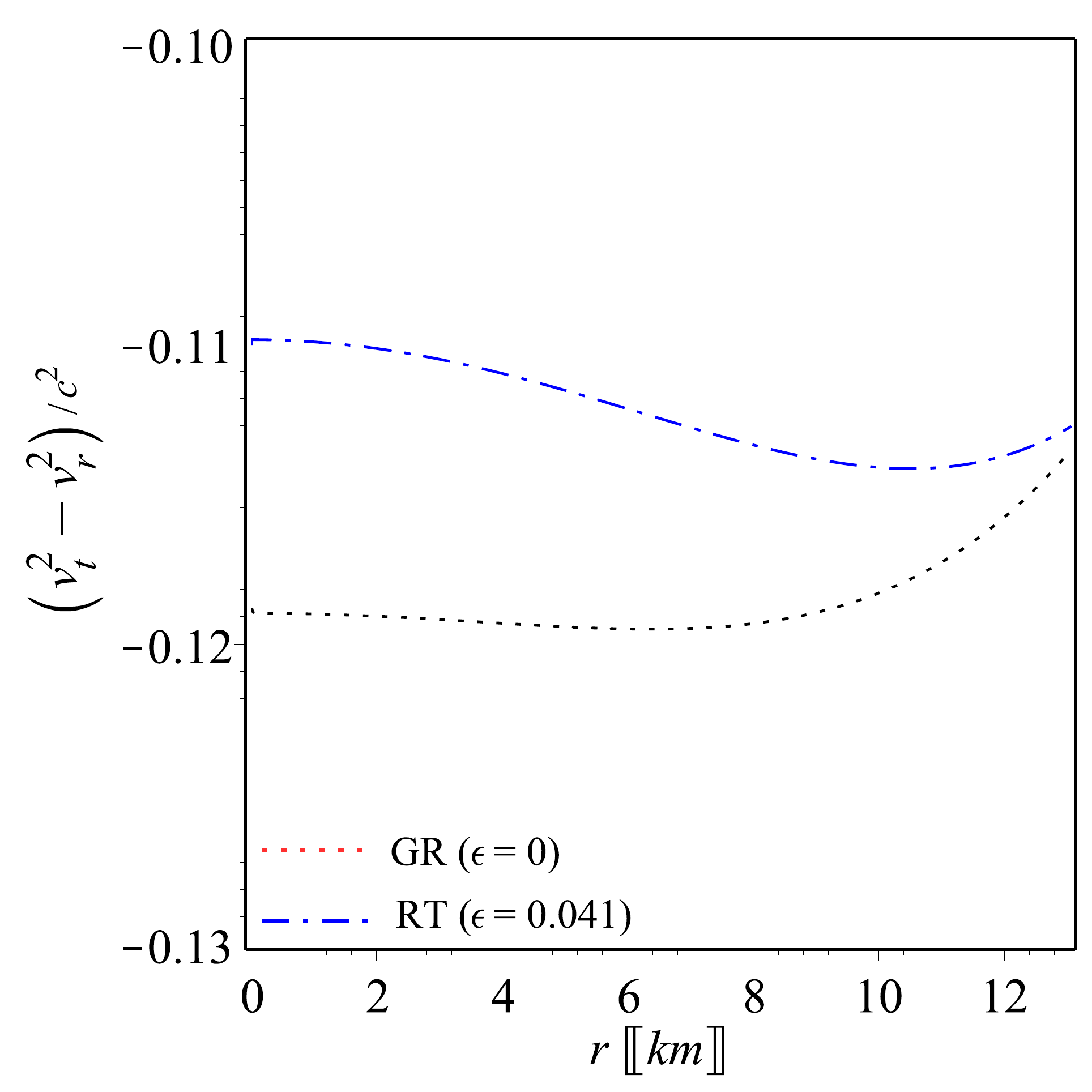}}
\caption{The radial and tangential sound speeds \eqref{eq:sound_speed} of the pulsar PSR J0740+6620. The plots confirm that the causality and stability conditions (\textbf{ix}) and (\textbf{x}) are satisfied.}
\label{Fig:Stability}
\end{figure*}
We define the square of the sound speed in radial and tangential directions,
\begin{equation}\label{eq:sound_speed}
  v_r^2 = \frac{dp_r}{d\rho}=\frac{\tilde{p}'_r}{\bar{\rho}'}, \quad v_t^2 = \frac{dp_t}{d\rho}=\frac{\tilde{p}'_t}{\bar{\rho}'},
\end{equation}
where the density and pressures radial gradients are given by Eqs. \eqref{eq:dens_grad}--\eqref{eq:pt_grad}.\\

\noindent Condition (\textbf{ix}): The stellar structure should be causal, i.e. satisfies the causality condition: sound speeds should be positive and smaller than unity everywhere inside the star ($0\leq v_r/c\leq 1$, $0\leq v_t/c \leq 1$), and monotonically decrease toward the boundary ($v'_r{^2}<0$, $v'_t{^2}<0$).

\noindent Condition (\textbf{x}): The stellar structure should be stable, i.e. satisfies the stability condition $-1< (v_t^2-v_r^2)/c^2 < 0$ everywhere inside the star \cite{HERRERa1992206}.

\textit{Obviously the causality and stability conditions} (\textbf{ix}) and (\textbf{x}) \textit{are satisfied for the pulsar PSR J0740+6620 as obtained by Fig. \ref{Fig:Stability}\subref{fig:vr}--\subref{fig:vt-vr}}.

Although the sound speed fulfills the causality and stability conditions for the pulsar PSR J0740+6620, the GR prediction of the sound speed within the region $r \lesssim 10.14$ km from the center exceeds the conjectured conformal upper bound on the sound speed $v_r^2=c_s^2\leq c^2/3$. This violation is strongly suggested for NS with radii less than $\sim 11.8$ km \citep{Bedaque:2014sqa}. Also, the violation of the conformal upper limit of the sound speed has been obtained in other models when hadronic EoS is assumed \citep{Cherman:2009tw,Landry:2020vaw} or when a non-parametric EoS approach based on Gaussian processes is applied to the pulsar PSR J0740+6620 using the X-ray NICER+XMM observations \citep{Legred:2021hdx}. On the contrary, as seen in Fig. \ref{Fig:Stability}\subref{fig:vr}, the conformal bound on the sound speed for the pulsar PSR J0740+6620 is not violated in RT from the core to the surface.
%%%%%%%%%%%%%%%%%%%%%%%%%%%%%%%%%%%%%%%%%%%%%%%%%%%%%%%%%%%%%%%%%%
\subsection{Adiabatic indices and hydrodynamic equilibrium}
We apply two further tests to examine the stability of the obtained model within Ratall gravity. First, we investigate the relativistic adiabatic indices of a spherically symmetric, which define the ratio of two specific heats \citep{1964ApJ...140..417C,1989A&A...221....4M,10.1093/mnras/265.3.533}.
\begin{equation}\label{eq:adiabatic}
\gamma=\frac{4}{3}\left(1+\frac{F_a}{2 |p'_r|}\right)_{max},\,
\Gamma_r=\frac{\rho c^2+p_r}{p_r} v_r^2,\,
\Gamma_t=\frac{\rho c^2+p_t}{p_t} v_t^2.
\end{equation}
For anisotropic fluid, the sphere is in a neutral (stable) equilibrium where the adiabatic indices $\Gamma=\gamma$ ($\Gamma>\gamma$) \citep[see][]{10.1093/mnras/265.3.533}. Clearly the adiabatic indices reduce to the isotropic sphere case where $\gamma=4/3$ and $\Gamma_r=\Gamma_t$ \citep{1975A&A....38...51H}. Second, we investigate the validity of TOV equation \citep{PhysRev.55.364,PhysRev.55.374,PoncedeLeon1993} by assuming the sphere is in a hydrostatic equilibrium everywhere inside the star where the acting forces neutralize each other. The modified version of TOV equation according to the newly introduced Rastall force $F_R$ can be written as
\begin{equation}\label{eq:RS_TOV}
F_a+F_g+F_h+F_R=0\,,
\end{equation}
where $F_g$ and $F_h$ denote the gravitational and the hydrostatic forces. Here we have
\begin{eqnarray}\label{eq:Forces}
% \nonumber to remove numbering (before each equation)
  F_a &=& 2\Delta/r ,\quad
  F_g = -\frac{M_g}{r}(\rho c^2+p_r)e^{\delta/2} ,\nonumber\\
  F_h &=& -p'_r ,\quad
  F_R = -\frac{\epsilon}{1-4\epsilon}(c^2 \rho'-p'_r-2p'_t),
\end{eqnarray}
where $\delta\equiv \delta(r)=\alpha-\beta$ and the mass (energy) $M_g$ for an isolated time-independent systems within 3-space $V$ ($t=$ constant) is given by Tolman mass formula \citep{PhysRev.35.896,book:293729}
\begin{eqnarray}\label{eq:grav_mass}
M_g(r)&=&{\int_V}\Big(-\mathfrak{T}{^t}{_t}+\mathfrak{T}{^r}{_r}+\mathfrak{T}{^\theta}{_\theta}+\mathfrak{T}{^\phi}{_\phi}\Big)\sqrt{-g}\,dV\nonumber\\
&=&\frac{(e^{\alpha/2})'}{e^{\alpha}} e^{\beta/2} r =\frac{\alpha'}{2} r e^{-\delta/2}.
\end{eqnarray}
This reads the gravitational force $F_g=-\frac{a_0 r}{R^2}(\rho c^2+p_r)$. Now we state explicitly the stability conditions related to the relativistic adiabatic indices and the modified TOV equation.
\begin{figure}
\centering
{\includegraphics[scale=0.28]{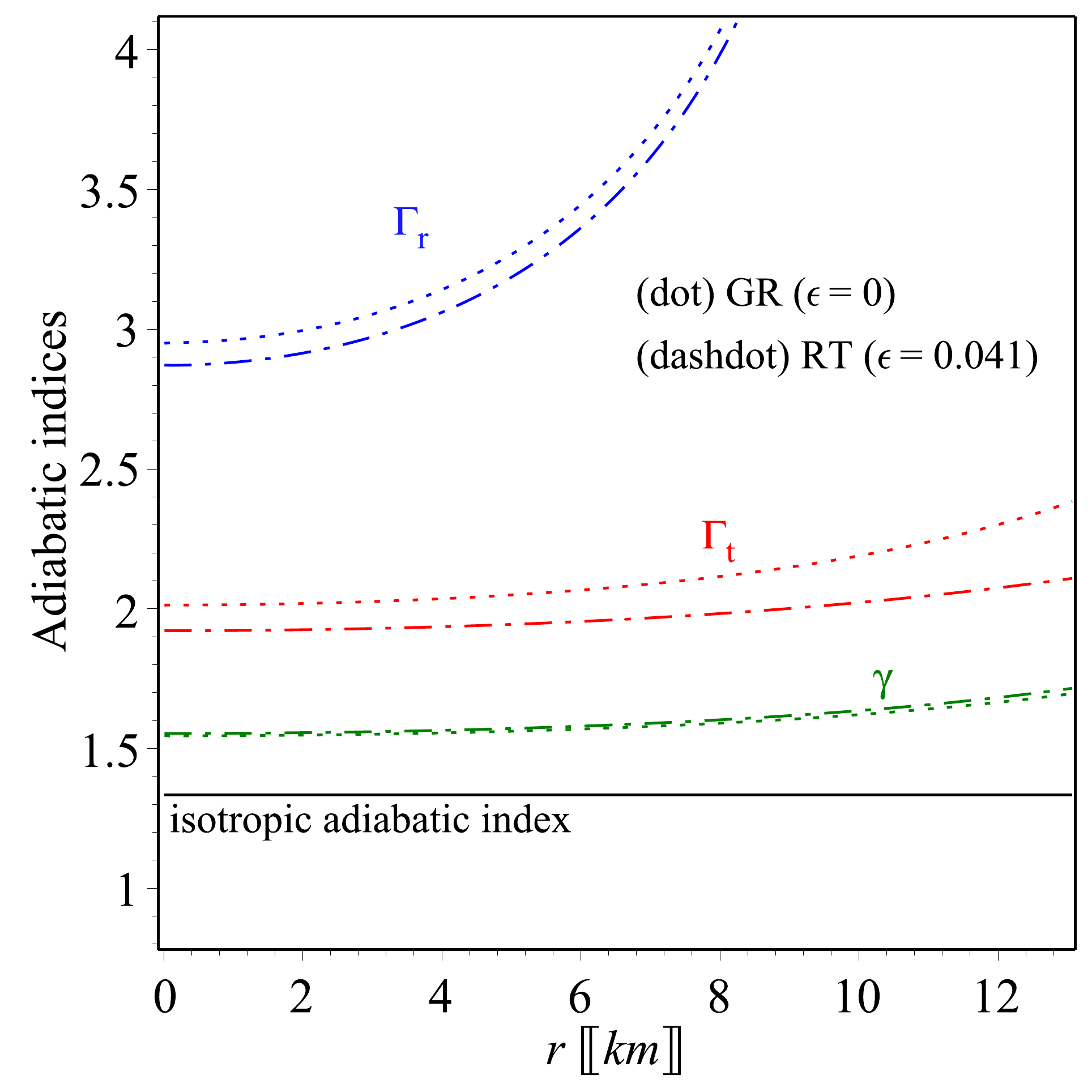}}
\caption{The adiabatic indices \eqref{eq:adiabatic} of the pulsar PSR J0740+6620. The plots confirm that the stability conditions (\textbf{xi}) are satisfied within the pulsar. Dotted (dash-dotted) curves represent GR (RT) adiabatic indices.}
\label{Fig:Adiab}
\end{figure}
\begin{figure}
\centering
\label{fig:GRTOV}\includegraphics[scale=0.28]{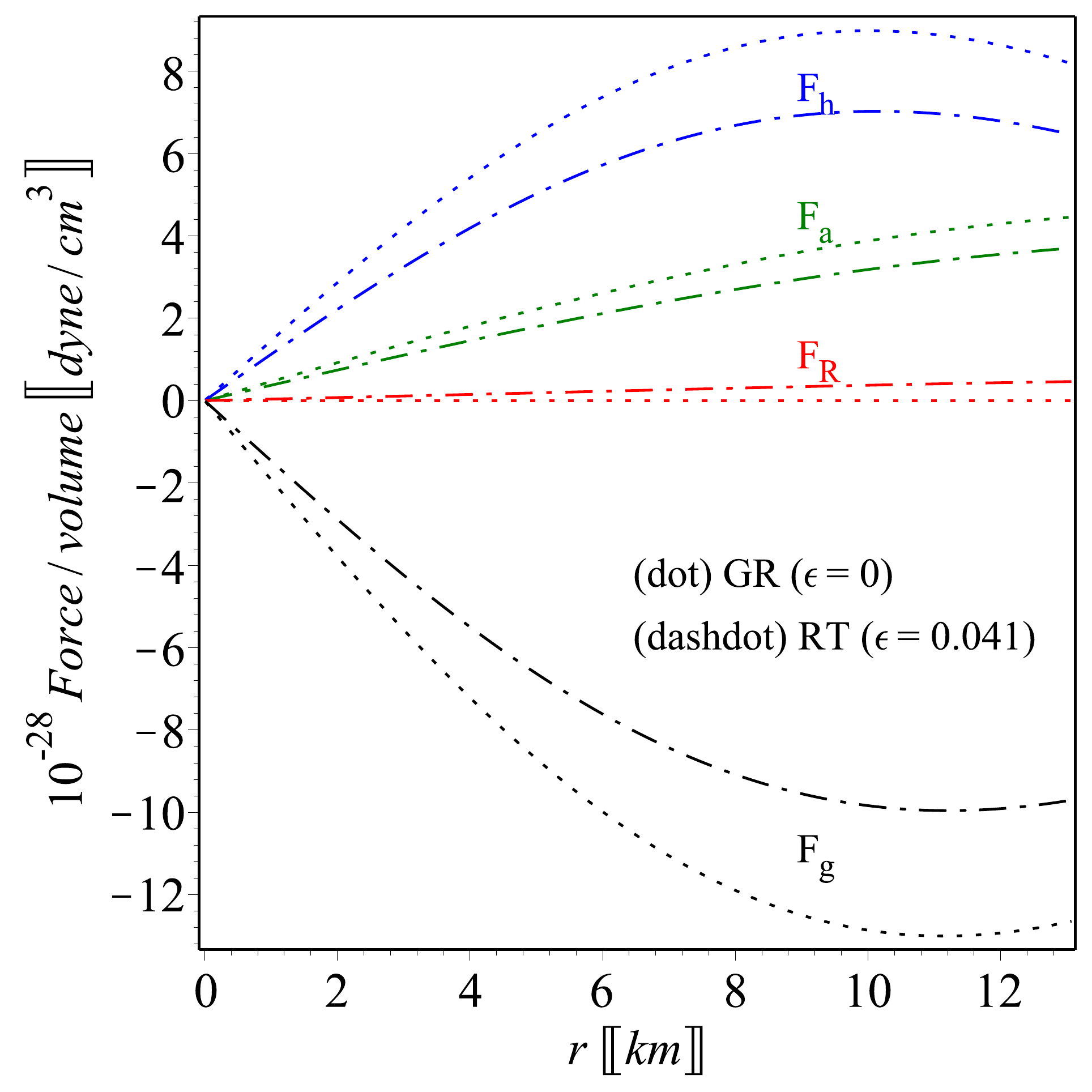}
%\subfigure[~GR]{\label{fig:GRTOV}\includegraphics[scale=0.25]{Fig8a.pdf}}
%\subfigure[~RT]{\label{fig:RTTOV}\includegraphics[scale=0.25]{Fig8b.pdf}}
\caption{The forces \eqref{eq:Forces} of the modified TOV equation \eqref{eq:RS_TOV} of the pulsar PSR J0740+6620. Dotted (dash-dotted) curves represent GR (RT) forces. The plots confirm that the hydrodynamic stability condition (\textbf{xii}) is satisfied within the pulsar.}
\label{Fig:TOV}
\end{figure}

\noindent Condition (\textbf{xi}): The anisotropic stellar model is in a stable equilibrium whereas the adiabatic indices satisfy the conditions $\Gamma_r> \gamma$ and $\Gamma_t> \gamma$ everywhere inside the star. Recalling the density and pressures Eqs. \eqref{eq:Feqs2} and their radial gradients \eqref{eq:dens_grad}--\eqref{eq:pt_grad} we plot the adiabatic indices \eqref{eq:adiabatic} of the pulsar PSR J0740+6620 for GR and RT cases in Fig. \ref{Fig:Adiab}. Noting that $\gamma=4/3$ for isotropic sphere.

\textit{Obviously the adiabatic indices stability conditions} (\textbf{xi}) \textit{are satisfied for the pulsar PSR J0740+6620 as obtained by Fig. \ref{Fig:Adiab}}.

\noindent Condition (\textbf{xii}): The anisotropic stellar model is in a hydrodynamic equilibrium whereas the forces satisfy The modified TOV equation \eqref{eq:RS_TOV}. We use Eqs. \eqref{eq:Feqs2} and \eqref{eq:dens_grad}--\eqref{eq:pt_grad} to evaluate the forces \eqref{eq:Forces} which are plotted in Fig. \ref{Fig:TOV} for GR and RT cases. The plots show that the negative gravitational force compensates (the positive forces) for a hydrodynamics equilibrium as required for a stable configuration. The role of Rastall force to resize compact objects will be discussed later in Sec. \ref{Sec:MR-reln}.

\textit{Obviously the hydrodynamic equilibrium condition} (\textbf{xii}) \textit{is satisfied for the pulsar PSR J0740+6620 as obtained by Fig. \ref{Fig:TOV}}.
%%%%%%%%%%%%%%%%%%%%%%%%%%%%%%%%%%%%%%%%%%%%%%%%%
\subsection{Neutron core, sound speed and EoS}

Using the X-ray NICER+XMM observational data of the pulsar PSR J0740+6620, we obtained a reasonable value of Rastall parameter $\epsilon=0.034$ which characterizes the coupling between matter and geometry in curved spacetime. Accordingly, we calculate the surface density which has been found as $\rho_R \approx 3.2\times 10^{14}$ g/cm$^{3}$ while at the center the density increases up to $\rho_\text{core} \approx 5.6\times 10^{14}$ g/cm$^3$ which is only $\sim 2.1$ times the nuclear saturation density $\rho_\text{nuc}=2.7\times 10^{14}$ g/cm$^{3}$. We note that the core density in RT with $\epsilon>0$ is less than the corresponding value in GR ($\epsilon=0$), since within a given radius the estimated mass by Rastall gravity is less than Einsteins gravity as seen in Fig. \ref{Fig:Mass}. The central density value suggests that the PSR J0740+6620 core is made of neutrons.
\begin{figure}[th!]
\centering
\subfigure[~Radial EoS]{\label{fig:REoS}\includegraphics[scale=0.3]{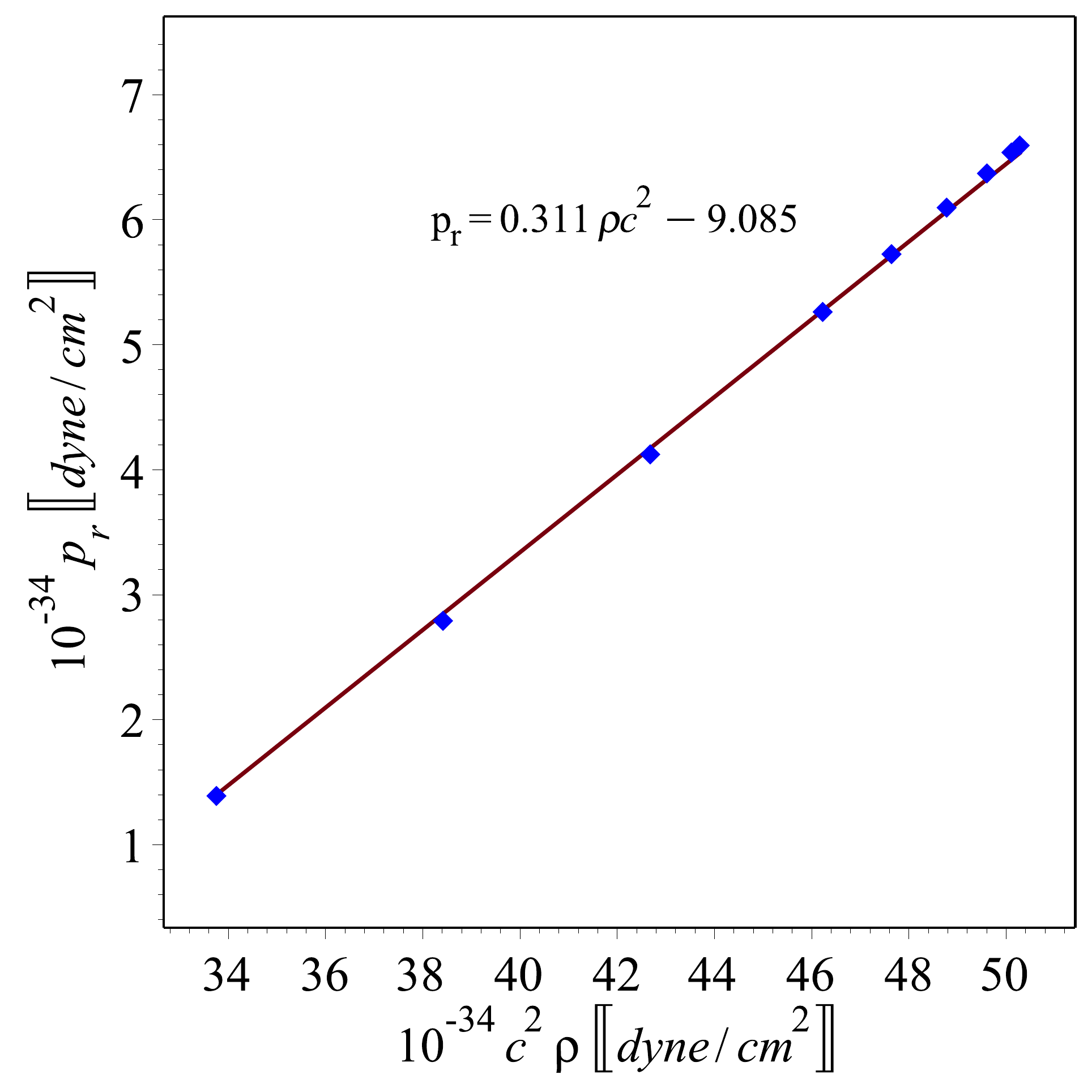}}
\subfigure[~Tangential EoS]{\label{fig:TEoS}\includegraphics[scale=.3]{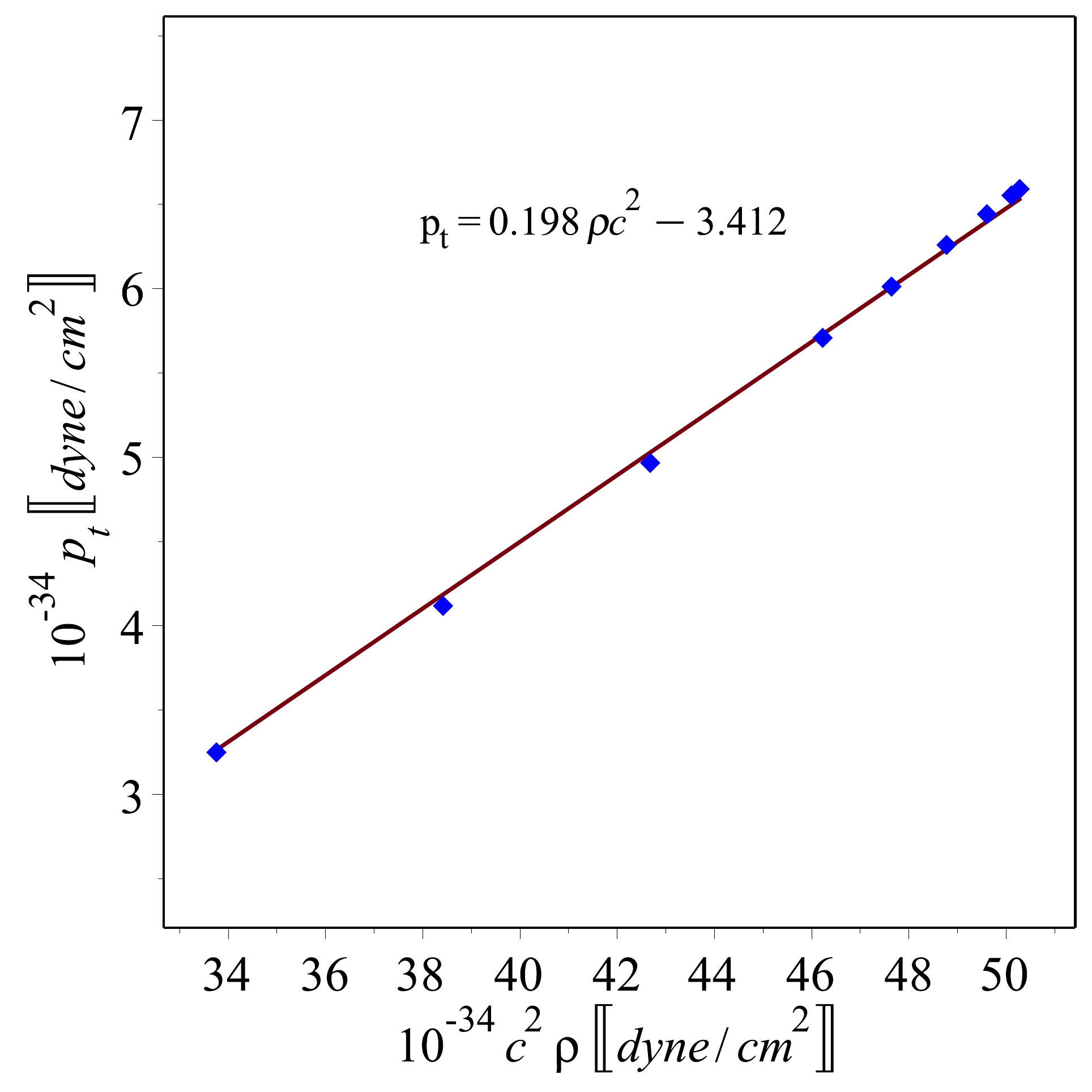}}
\caption{The pressures-density relation of the pulsar PSR J0740+6620 is in a good agreement with the linear EoS with bag constant whereas the slopes $dp_r/d\rho=0.312 c^2$ and $dp_t/d\rho=0.199 c^2$ evidently confirm that the NS matters fulfill the conjectured conformal bound on the sound speed $c_s^2\leq c^2/3$ everywhere inside the star.}
\label{Fig:EoS}
\end{figure}

It is to be noted that the NS with $\sim 2 M_\odot$ with hadronic EoS strongly violates the conjectured conformal bound on the sound speed $c_s^2\leq c^2/3$ even at low densities \cite{Bedaque:2014sqa}. This may point out a strongly interacting and nonconformal matter at the NS core \cite{Cherman:2009tw,Landry:2020vaw}. A non-parametric EoS approach based on Gaussian processes has been applied to investigate the implications of the X-ray NICER+XMM observations of the pulsar PSR J0740+6620 in particular on high density matter \cite{Legred:2021hdx}, it has been concluded that the conformal sound speed is strongly violated at the core whereas $c_s^2=0.75 c^2$. Interestingly, for the model at hand without imposing EoS, we find that the maximum squared sound speeds at the NS core $v^2_{r} \lesssim 0.332 c^2$ (radial direction) and $v^2_{t} \lesssim 0.222 c^2$ (tangential direction), which satisfy the conjectured conformal bound on the sound speed $c_s^2\leq c^2/3$ at the core and also everywhere inside the star, see Fig. \ref{Fig:Stability}. This unlike the GR case whereas $c_s^2\approx 0.37 c^2$.

Although we do not preassume a particular EoS, we find that the pressure-density relation according to the currant model is in a good agreement with the linear EoS with bag constant whereas $p_r(\rho)=0.311 \rho c^2-9.085$ and $p_t(\rho)=0.198 \rho c^2-3.412$ as seen in Figs. \ref{Fig:EoS}. Notably the slopes of the fitted lines $dp_r/d\rho=0.311 c^2$ and $dp_t/d\rho=0.198 c^2$ are in agreement with the obtained squared sound speeds in radial and tangential directions which also confirm the fulfillment of the conjectured conformal bound on the sound speed $c_s^2\leq c^2/3$ everywhere inside the star. We also note that the present anisotropic NS model obtains an EoS which is not too stiff as usually found for NS models, but in agreement with the EoS as expected from gravitational wave signals where no clear evidence on a tidal deformation in the observed GW patterns.
%%%%%%%%%%%%%%%%%%%%%%%%%%%%%%%%%%%%%%%%%%%%%%%%%%%%%%%%%%%%%%%%%%%%%%%%%%%%%%%%%%%%%%%%
\section{Confronting the model with more pulsar data}\label{Sec:pulsars}
In the previous section we confronted the model with the X-ray NICER+XMM observational data of the pulsar PSR J0740+6620 constraining Rastall parameter to a value of $\epsilon=0.041$. We extend our investigation to include more pulsars with different types of observations to examine the applicability of the model to a wide range of compact stars.
\begin{table*}
\caption{Observed mass-radius of twenty pulsars and the corresponding model parameters ($\epsilon=0.06$).}
\label{Table1}
\begin{tabular*}{\textwidth}{@{\extracolsep{\fill}}llcccccc@{}}
\hline\hline
\multicolumn{1}{c}{Pulsar} & \multicolumn{1}{c}{Ref} & observed mass & observed radius & estimated mass & \multicolumn{1}{c}{$a_0$}
& \multicolumn{1}{c}{$a_1$}& \multicolumn{1}{c}{$a_2$}\\
   & &  ($M_{\odot}$) &   [{km}] &  ($M_{\odot}$) &      &      &    \\
\hline
\multicolumn{8}{c}{High-mass X-ray binaries}\\
\hline
Her X-1         &\cite{Abubekerov_2008}        &  $0.85\pm 0.15$    &  $8.1\pm 0.41$   & $0.767$&  $0.199$    & $-0.570$    & $0.371$     \\
4U 1538-52      &\cite{Gangopadhyay:2013gha}   &  $0.87\pm 0.07$    &  $7.866\pm 0.21$ & $0.804$&  $0.223$    & $-0.631$    & $0.407$     \\
LMC X-4         &\cite{Rawls:2011jw}           &  $1.04\pm 0.09$    &  $8.301\pm 0.2$  & $0.954$&  $0.269$    & $-0.742$    & $0.472$     \\
Cen X-3         &\cite{Naik:2011qc}            &  $1.49\pm 0.49$    &  $9.178\pm 0.13$ & $1.344$&  $0.417$    & $-1.070$    & $0.653$     \\
\hline
\multicolumn{8}{c}{Low-mass X-ray binaries (quiescence/thermonuclear bursts)}\\
\hline
EXO 1785-248    &\cite{Ozel:2008kb}            &  $1.3\pm 0.2$      &  $8.849\pm 0.4$  & $1.172$&  $0.344$    & $-0.914$    & $0.569$     \\
M13             &\cite{Webb:2007tc}            &  $1.38\pm 0.2$     &  $9.95\pm 0.27$  & $1.245$&  $0.311$    & $-0.838$    & $0.527$     \\
X7              &\cite{Bogdanov:2016nle}       &  $1.1\pm 0.35$     &  $12$            & $1.073$&  $0.183$    & $-0.530$    & $0.347$     \\
4U 1820-30      &\cite{Ozel:2015fia}           &  $1.46\pm 0.2$     &  $11.1\pm 1.8$   & $1.317$&  $0.284$    & $-0.775$    & $0.492$     \\
4U 1608-52      &\cite{1996IAUC.6331....1M}    &  $1.57\pm 0.3$     &  $9.8\pm 1.8$    & $1.470$&  $0.438$    & $-1.114$    & $0.676$     \\
KS 1731-260     &\cite{Ozel:2008kb}            &  $1.61\pm 0.37$    &  $10\pm 2.2$     & $1.452$&  $0.410$    & $-1.055$    & $0.646$     \\
EXO 1745-268    &\cite{Ozel:2008kb}            &  $1.65\pm 0.25$    &  $10.5\pm 1.8$   & $1.488$&  $0.391$    & $-1.015$    & $0.624$     \\
4U 1724-207     &\cite{Ozel:2008kb}            &  $1.81\pm 0.27$    &  $12.2\pm 1.4$   & $1.632$&  $0.351$    & $-0.928$    & $0.577$     \\
SAX J1748.9-2021&\cite{Ozel:2008kb}            &  $1.81\pm 0.3$     &  $11.7\pm 1.7$   & $1.632$&  $0.379$    & $-0.990$    & $0.611$     \\
\hline
\multicolumn{8}{c}{Millisecond Pulsars}\\
\hline
PSR J0030+0451  &\cite{Raaijmakers:2019qny}    &  $1.34\pm 0.16$    &  $12.71\pm 1.19$ & $1.266$&  $0.215$    & $-0.610$    & $0.395$     \\
PSR J0030+0451  &\cite{Miller:2019cac}         &  $1.44\pm 0.16$    &  $13.02\pm 1.24$ & $1.313$&  $0.219$    & $-0.620$    & $0.401$     \\
PSR J0437-4715  &\cite{Reardon:2015kba}        &  $1.44\pm 0.07$    &  $13.6\pm 0.9$   & $1.387$&  $0.223$    & $-0.629$    & $0.407$     \\
                &\& \cite{Gonzalez-Caniulef:2019wzi}        &      &     &  &      &     &       \\
PSR J1614-2230  &\cite{NANOGrav:2017wvv}       &  $1.908\pm 0.016$  &  $13\pm 2$       & $1.894$&  $0.413$    & $-1.06$     & $0.649$     \\
PSR J0348+0432  &\cite{Antoniadis:2013pzd}     &  $2.01\pm 0.04$    &  $13\pm 2$       & $1.975$&  $0.387$    & $-1.008$    & $0.620$     \\
\hline
\multicolumn{8}{c}{Gravitational-wave Signals}\\
\hline
LIGO-Virgo      &\cite{LIGOScientific:2020zkf} &  $1.4$             &  $12.9\pm 0.8$   & $1.400$&  $0.246$    & $-0.685$    & $0.440$     \\
GW170817-1      &\cite{LIGOScientific:2018cki} &  $1.45\pm 0.09$    &  $11.9\pm 1.4$   & $1.263$&  $0.209$    & $-0.596$    & $0.387$     \\
GW170817-2      &\cite{LIGOScientific:2018cki} &  $1.27\pm 0.09$    &  $11.9\pm 1.4$   & $1.210$&  $0.222$    & $-0.627$    & $0.405$     \\
\hline
\end{tabular*}
\end{table*}

In Table \ref{Table1} we list the pulsars according to different types observations with their corresponding constraints on masses and radii. We include: (i) four high mass X-ray binaries; (ii) nine low mass X-ray binaries, where the mass-radius constraints are given by using spectral analysis techniques for two quiescent low X-ray binaries (M13 and X7), while others are characterized by Type I X-ray bursts thermonuclear explosions on the neutron star surface; (iii) four MSPs including two of the most massive pulsars PSR J0348+0432 and PSR J1614-2230 where the mass-radius constraints are obtained by measuring the arrival time of pulses, we further include the recent NICER observational constraints on the two pulsars PSR J0437-4715 and PSR J0030+0451; (iv) three gravitational-wave signals including the first detected NS-NS merger GW170817 in addition to LIGO-Virgo constraints on NS radius from GW170817+GW190814. Notably the companion of the GW190814 binary is the first detected compact object in the low mass gap $M=2.6 M_\odot$. In addition, we estimate the values of the pulsars' masses (corresponding to the observed/estimated mean value of $R$) with the corresponding values of the model parameters $a_0$, $a_1$ and $a_2$. As seen from the table, the estimated masses are compatible with the corresponding observed values.

As mentioned in the introduction that NICER observations of PSR J0030+0451 and PSR J0740+6620 represents a challenge for squeezable models, since the latter has much more mass than the former while both are having almost same size. We note that the present model estimates ($M= 1.266 M_\odot, R=12.71$ km)  and ($M= 1.313 M_\odot, R=13.02$ km) for the pulsar PSR J0030+0451 in agreement with its NICER observational constraints as reported by \cite{Raaijmakers:2019qny} and \cite{Miller:2019cac} respectively. At the same time it estimates ($M= 1.96 M_\odot, R=13.04$ km) for the pulsar PSR J0740+6620 in agreement with NICER+XMM measurements \citep{Legred:2021hdx}. This is in agreement with our claim that when a pulsar gains more mass and subsequently higher density, the induced anisotropic force becomes relevant to hold the NS size.

We further examine the stability of the model according to the criteria given in the previous section. We list the physical quantities of the most interest in Table \ref{Table2}. The results show the capability of the model to predict stable compact stellar structure compatible with observations. Interestingly, for all categories the sound speeds as obtained by this model are in agreement with the conformal bound $c^2_s \leq c^2/3$. It is to be noted that for the low mass X-ray binaries 4U 1608-52 and KS 1731-260, in particular, the general relativistic version of this study predicted $c^2_s=0.34$ at the center above the conformal bound as obtained by \cite{Roupas:2020mvs}. Similar to our previous discussion on the sound speed of the pulsar PSR J0740+6620, the results of Table \ref{Table2} confirm that the sound speed violation is avoided in the present study. This shows that the possible role of the geometry-matter coupling to hold this upper bound  everywhere inside the compact object.
\begin{table*}
\caption{Calculated physical quantities of the most interest.}
\label{Table2}
\begin{tabular*}{\textwidth}{@{\extracolsep{\fill}}l|cc|cc|cc|cc|cc@{\extracolsep{\fill}}}
\hline\hline
\multicolumn{1}{c|}{Pulsar}                              &\multicolumn{2}{c|}{$\rho$ [g/cm$^3$]} &   \multicolumn{2}{c|}{$v_r^2/c^2$}  &   \multicolumn{2}{c|}{$v_t^2/c^2$}  & \multicolumn{2}{c|}{DEC [$dyne/cm^2$]} & \multicolumn{1}{c}{$Z_R$}\\ \cline{2-10}
                                        &\multicolumn{1}{c}{center}            &        surface   &   center        &    surface       &  center          &  surface        &  center     & surface       & {}\\
\hline
\multicolumn{10}{c}{High-mass X-ray binaries}\\
\hline
Her X-1             &7.89$\times10^{14}$     &5.80$\times10^{14}$  &  0.255   &   0.228     &  0.153 & 0.118 & 5.72$\times10^{35}$ & 4.80$\times10^{35}$ & 0.204  \\
4U 1538-52          &9.19$\times10^{14}$     &6.55$\times10^{14}$  &  0.264   &   0.233     &  0.161 & 0.122 & 6.49$\times10^{35}$ & 5.38$\times10^{35}$ & 0.226  \\
LMC X-4             &9.59$\times10^{14}$     &6.49$\times10^{14}$  &  0.279   &   0.242     &  0.175 & 0.131 & 6.42$\times10^{35}$ & 5.23$\times10^{35}$ & 0.266  \\
Cen X-3             &1.10$\times10^{15}$     &6.39$\times10^{14}$  &  0.326   &   0.271     &  0.217 & 0.158 & 6.12$\times10^{35}$ & 4.83$\times10^{35}$ & 0.386  \\
\hline
\multicolumn{10}{c}{Low-mass X-ray binaries (quiescence/thermonuclear bursts)}\\
\hline
EXO 1785-248        &1.02$\times10^{15}$     &6.38$\times10^{14}$  &  0.303   &   0.257     &  0.197 & 0.145 & 6.25$\times10^{35}$ & 4.98$\times10^{35}$ & 0.329  \\
M13                 &7.46$\times10^{14}$     &4.83$\times10^{14}$  &  0.292   &   0.251     &  0.188 & 0.139 & 4.76$\times10^{35}$ & 3.82$\times10^{35}$ & 0.302  \\
X7                  &3.36$\times10^{14}$     &2.51$\times10^{14}$  &  0.250   &   0.224     &  0.147 & 0.114 & 2.48$\times10^{35}$ & 2.10$\times10^{35}$ & 0.189  \\
4U 1820-30          &5.59$\times10^{14}$     &3.72$\times10^{14}$  &  0.284   &   0.245     &  0.180 & 0.134 & 3.68$\times10^{35}$ & 2.98$\times10^{35}$ & 0.279  \\
4U 1608-52          &9.91$\times10^{14}$     &5.70$\times10^{14}$  &  0.332   &   0.275     &  0.222 & 0.162 & 5.41$\times10^{35}$ & 4.27$\times10^{35}$ & 0.402  \\
KS 1731-260         &9.08$\times10^{14}$     &5.35$\times10^{14}$  &  0.324   &   0.270     &  0.215 & 0.157 & 5.13$\times10^{35}$ & 4.06$\times10^{35}$ & 0.381  \\
EXO 1745-268        &7.95$\times10^{14}$     &4.77$\times10^{14}$  &  0.318   &   0.266     &  0.210 & 0.154 & 4.61$\times10^{35}$ & 3.65$\times10^{35}$ & 0.366  \\
4U 1724-207         &5.44$\times10^{14}$     &3.38$\times10^{14}$  &  0.305   &   0.286     &  0.199 & 0.146 & 3.31$\times10^{35}$ & 2.63$\times10^{35}$ & 0.334  \\
SAX J1748.9-2021    &6.27$\times10^{14}$     &3.80$\times10^{14}$  &  0.314   &   0.264     &  0.207 & 0.151 & 3.68$\times10^{35}$ & 2.92$\times10^{35}$ & 0.357  \\
\hline
\multicolumn{10}{c}{Millisecond Pulsars}\\
\hline
PSR J0030+0451      &3.41$\times10^{14}$     &2.46$\times10^{14}$  &  0.260   &   0.231     &  0.158 & 0.121 & 2.43$\times10^{35}$ & 2.03$\times10^{35}$ & 0.218  \\
PSR J0030+0451      &3.30$\times10^{14}$     &2.37$\times10^{14}$  &  0.262   &   0.231     &  0.159 & 0.121 & 2.34$\times10^{35}$ & 1.95$\times10^{35}$ & 0.222  \\
PSR J0437-4715      &3.07$\times10^{14}$     &2.19$\times10^{14}$  &  0.263   &   0.233     &  0.160 & 0.122 & 2.17$\times10^{35}$ & 1.80$\times10^{35}$ & 0.225  \\
PSR J1614-2230      &5.39$\times10^{14}$     &3.17$\times10^{14}$  &  0.325   &   0.270     &  0.216 & 0.158 & 3.04$\times10^{35}$ & 2.40$\times10^{35}$ & 0.383  \\
PSR J0348+0432      &4.45$\times10^{14}$     &2.67$\times10^{14}$  &  0.317   &   0.266     &  0.209 & 0.153 & 2.59$\times10^{35}$ & 2.05$\times10^{35}$ & 0.364 \\
\hline
\multicolumn{10}{c}{Gravitational-wave Signals}\\
\hline
LIGO-Virgo          &3.69$\times10^{14}$     &2.56$\times10^{14}$  &  0.271   &   0.238     &  0.168 & 0.126 & 2.54$\times10^{35}$ & 2.09$\times10^{35}$ & 0.246  \\
GW170817-1          &3.24$\times10^{14}$     &2.35$\times10^{14}$  &  0.259   &   0.230     &  0.156 & 0.120 & 2.32$\times10^{35}$ & 1.94$\times10^{35}$ & 0.213  \\
GW170817-2          &3.99$\times10^{14}$     &2.85$\times10^{14}$  &  0.263   &   0.233     &  0.160 & 0.122 & 2.82$\times10^{34}$ & 2.34$\times10^{34}$ & 0.225  \\
\hline
\end{tabular*}
\tablecomments{For the pulsars 4U 1608-52 and KS 1731-260, the sound speed at the core is in agreement with the conjectured conformal bound on the sound speed $c_s^2\leq c^2/3$, unlike the GR version \citep{Roupas:2020mvs}. This confirms the role of the matter-geometry coupling to avoid the violation of the upper conformal bound of the sound speed.}
\end{table*}
%
%%%%%%%%%%%%%%%%%%%%%%%%%%%%%%%%%%%%%%%%%%%%%% Section 5 %%%%%%%%%%%%%%%%%%%%%%%%%%%%%%%%%%%%%%%%%%%%%%%%%%%%%%%%%%%%%%%
\section{Maximum Compactness and Mass-Radius Relation}\label{Sec:MR-reln}

As is shown in the above sections the physical quantities of KB interior solution ($0 \leq x \leq 1$) are expressed in terms of Rastall and compactness parameters. On the other hand, observational constraints on masses and radii allow us to estimate Rastall parameter to be at most $\epsilon=0.041$. In this section we follow \citet{Roupas:2020mvs} utilizing the energy dominance condition\footnote{We note that any imposed constraints from the DEC for matter or effective fluids are identical as discussed in Sec. \ref{Sec:Energy-conditions}.}, namely $\rho c^2-p_r-2p_t \geq 0$ with $\epsilon=0.041$, to set an upper bound on the compactness value for the model at hand showing the role of Rastall parameter to support higher compactness values in comparison to GR predictions. In addition, we plot the mass-radius relation for some boundary conditions on the surface density compatible with nuclear density showing the maximum mass and radius for each case in comparison to GR predictions.
\subsection{Maximum compactness}
\begin{figure}
\centering
{\includegraphics[scale=0.3]{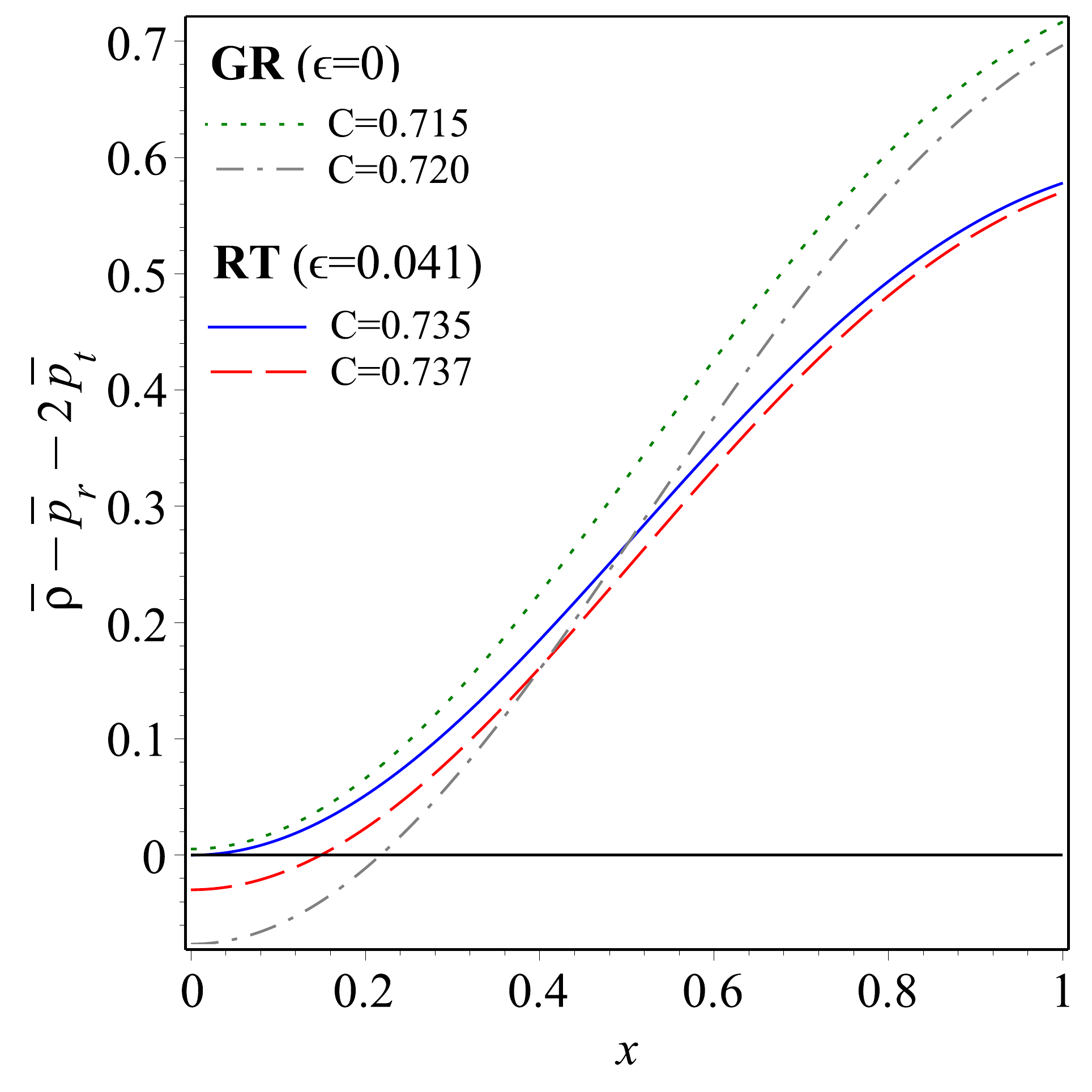}}
\caption{A comparison between GR ($\epsilon=0$) and RT ($\epsilon=0.041$) violation of the DEC $\bar{\rho}-\bar{p}_r-2\bar{p}_t \geq 0$ at higher compactness. We conclude that RT can support stable stellar configurations with higher compactness in comparison to GR.}
\label{Fig:Max_comp}
\end{figure}

Recalling the dimensionless energy density and pressures \eqref{eq:Feqs2} where the model parameters \{$a_0, a_1, a_2$\} are given by \eqref{eq:const}, we use the energy density dominance constraint to set an upper bound on the compactness which has been found to be $C_{|\epsilon = 0.041} \leq 0.735$ satisfying Buchdahl compactness bound $C\leq 8/9$ (for isotropic sphere \cite{PhysRev.116.1027}). In Fig. \ref{Fig:Max_comp}, we plot the energy density dominance patterns for GR ($\epsilon=0$) and RT ($\epsilon=0.041$) using different compactness values. The GR prediction of the maximum compactness is $C=0.715$ in agreement with \cite{Roupas:2020mvs}, however the contribution of the matter-geometry nonminimal coupling as assumed in RT allows $\sim 2\%$ maximum compactness higher in comparison to GR limits.

This result is in agreement with our earlier conclusion that RT with $\epsilon>0$ predicts larger pulsar size relative to GR for a given mass, see Fig. \ref{Fig:Mass}. Recalling TOV equation of the hydrodynamic equilibrium \eqref{eq:RS_TOV}, we find that the Rastall force contributes to oppose partially the gravitational force, see Fig. \ref{Fig:TOV}, to compensate for a larger size of a compact object of a given mass.

We further note that the obtained result of the upper bound on the compactness is in contrast to the suggestions that Buchdahl upper bound on the compactness can be violated in anisotropic stable compact stars \citep{Dev:2004ss, Bohmer2006}, also it sets a stronger restriction on the maximally allowed compactness than the corresponding limits due to surface redshift constraints \citep{Ivanov:2002xf}.

%%%%%%%%%%%%%%%%%%%%%%%%%%%%%%%%%%%%%%%%%%%%%%%%%%%%%%%%%%
\subsection{Mass-Radius diagram}
\begin{figure*}
\centering
\subfigure[~Compactness-Radius diagram]{\label{fig:Comp}\includegraphics[scale=0.28]{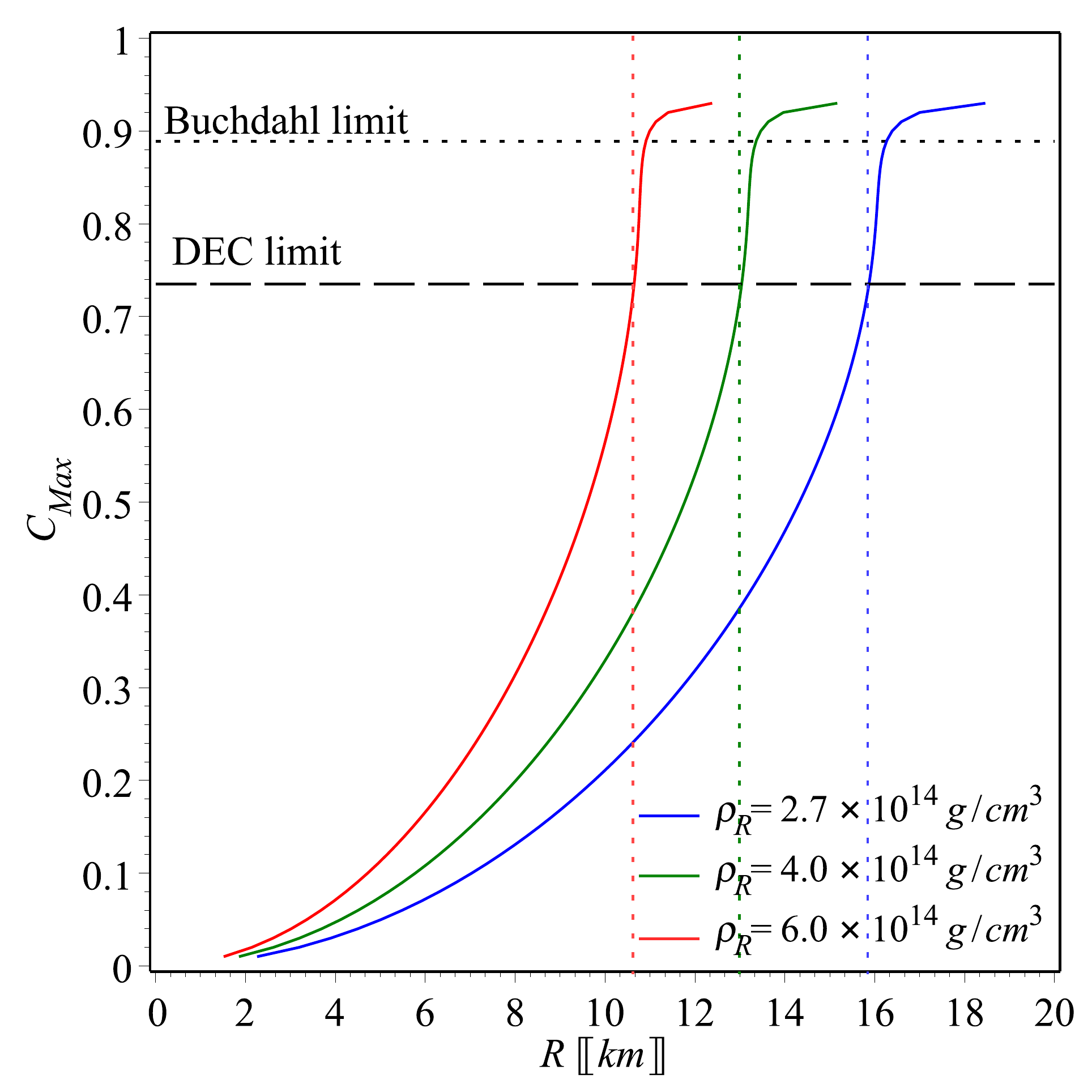}}
\subfigure[~Mass-Radius diagram]{\label{fig:MR}\includegraphics[scale=.28]{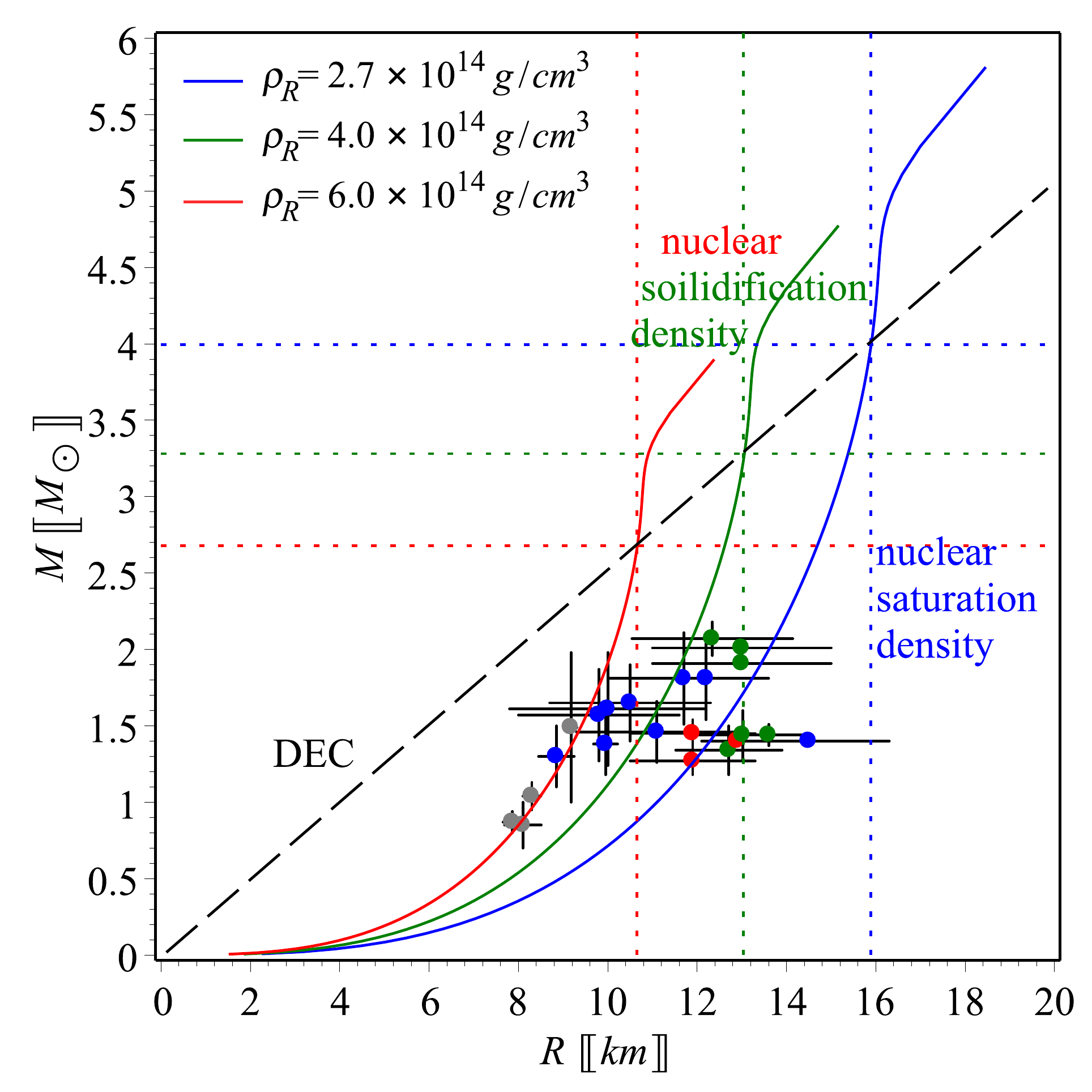}}
\subfigure[~LIGO-Virgo and NICER constraints]{\label{fig:NICER}\includegraphics[scale=.28]{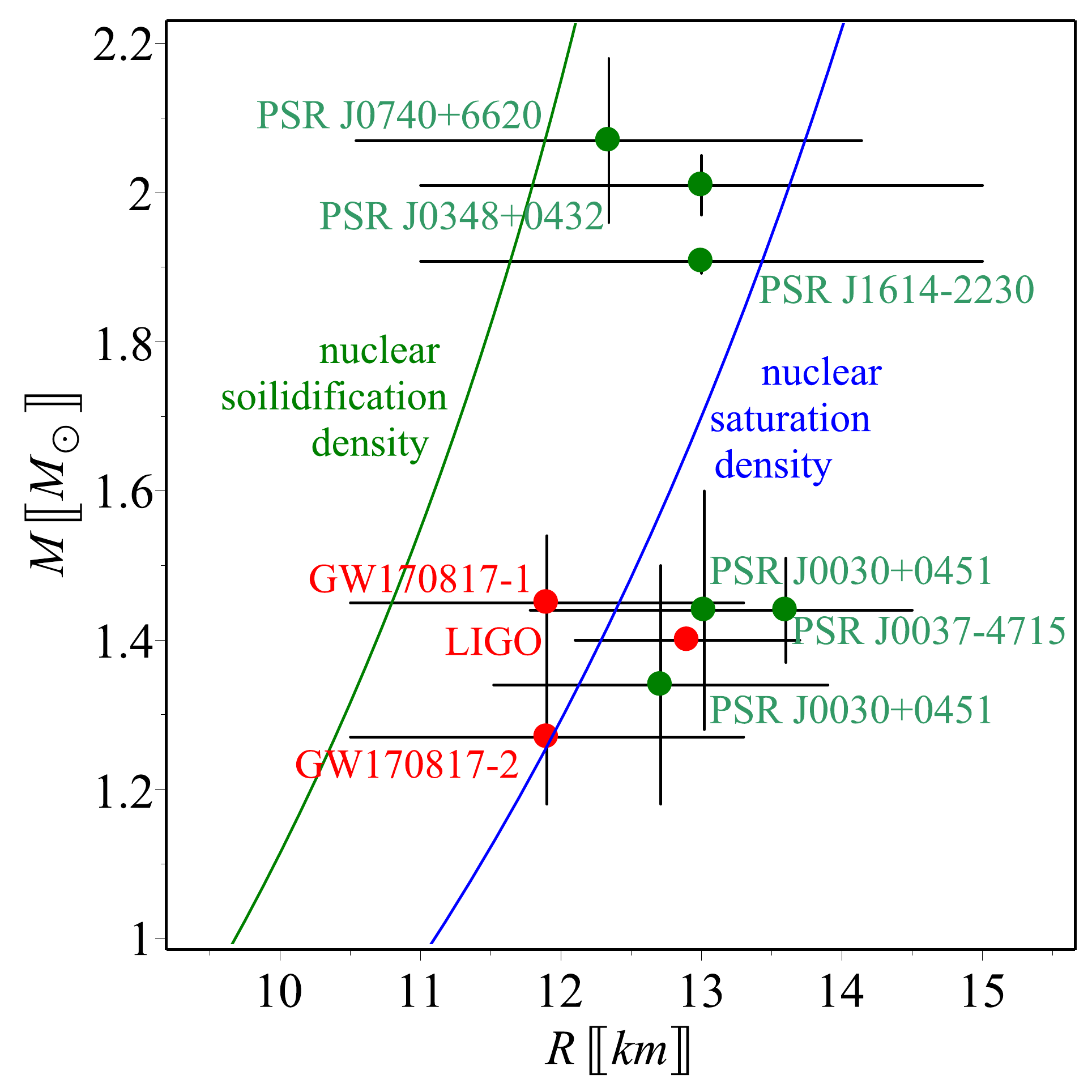}}
\caption{\subref{fig:Comp} The compactness-radius curves, at different surface densities, show that the maximum compactness is $C= 0.93$ just above Buchdahl limit $C=8/9$ (dotted horizontal line) for isotropic sphere. However, the strong energy condition sets a more strict constraint on the upper bound $C=0.735$ (dashed horizontal line) for an anisotropic compact object to be physically viable.
\subref{fig:MR} The mass-radius curves show that the DEC (diagonal dashed line) obtains an upper mass of $M\approx 4 M_\odot$ in the lower mass gap with radius $R\approx 16$ km whereas the surface density chosen to be at the saturation nuclear density $\rho_{\text{nuc}}=2.7\times 10^{14}$ g/cm$^3$. Solid gray circles represent high-mass X-ray binaries, solid blue circles represent low-mass X-ray binaries, solid green circles represent MSP and solid red circles represent GW signals as given from Table \ref{Table1},
\subref{fig:NICER} Close view on some particular pulsars with most interest. LIGO-Virgo constraints on the radius of a canonical NS, NICER constraints on the pulsar PSR J0030+0451 and NICER+XMM on the pulsar PSR J0740+6620 all are in agreement with the $\rho_\text{nuc}$-curve.}
\label{Fig:CompMR}
\end{figure*}

From Table \ref{Table2} one obtains surface densities $2\times 10^{14} \lesssim \rho_{R} \lesssim 7\times 10^{14}$ g/cm$^{3}$; and therefore we choose four surface densities to serve as boundary conditions as follows: $\rho_R=\rho_\text{nuc}=2.7\times 10^{14}$ g/cm$^3$ at nuclear saturation density, $\rho_R=4\times 10^{14}$ g/cm$^3$ and $\rho_R=6\times 10^{14}$ g/cm$^3$ to cover the nuclear solidification density. Then, for Rastall parameter $\epsilon=0.041$, we use the energy density equation \eqref{eq:Feqs2} to obtain a relation between $C$ and $R$ for each boundary condition on the surface density, i.e. $\rho(r=R)=\rho_R$. In Fig. \ref{Fig:CompMR}\subref{fig:Comp}, we plot the compactness-radius curves in correspondence to the chosen boundary density. The plots confirm that the maximum compactness is $C= 0.93$ just above Buchdahl limit $C=8/9$ (represented by a horizontal dotted line) for isotropic star. Remarkably the maximal compactness approaches a saturation-like pattern which reflects the capability of the NSs to gain more sizes with small increase of masses at this limit. However, as we obtained earlier in this section the energy dominance constraint sets an upper bound on the compactness $C=0.735$ (represented by a horizontal dashed line) for stable anisotropic configuration in KB spacetime, which consequently determines the maximal allowed radius for each boundary density as represented by the vertical dotted lines in Fig. \ref{Fig:CompMR}\subref{fig:Comp}.

Next we give the mass-radius curves for each case in addition to the corresponding observational data from Table \ref{Table1} as obtained by Fig. \ref{Fig:CompMR}\subref{fig:MR}. For the boundary density at the nuclear saturation density $\rho_\text{nuc}=2.7\times 10^{14}$ g/cm$^3$, applying the DEC constraint (represented by a diagonal dashed line), we calculate the maximum allowed mass $M_\text{max}=4 M_\odot$ at a maximum radius of $R_\text{max}=16$ km. Notably for the same boundary condition in KB spacetime the GR predicts a maximum mass $M_\text{max}=4.1 M_\odot$ at maximum radius $R_\text{max}=16.8$ km \citep{Roupas:2020mvs}. In comparison to GR predictions, RT with positive parameter $\epsilon=0.041$ predicts almost same mass within $\sim 0.8$ km smaller size than GR. Similarly, for the surface densities $\rho_R=4\times 10^{14}$ g/cm$^3$ and $\rho_R=6\times 10^{14}$ g/cm$^3$ which are compatible with the nuclear solidification density, we obtain the maximum masses and radii, ($M_\text{max}=3.28 M_\odot,\, R_\text{max}=13.18$ km) and ($M_\text{max}= 2.68 M_\odot,\, R_\text{max}=10.77$ km), respectively. Maximal radii (masses) allowed by the DEC (diagonal dashed line) are represented by vertical (horizontal) dotted lines for each boundary density on the mass-radius graph in Fig. \ref{Fig:CompMR}\subref{fig:MR}. Clearly, all pulsars lie in the physical region as allowed by the DEC. Remarkably, the present model can produce a NS within the mass gap $2.5-5 M_\odot$ which does not exclude that the companion of the binary GW190814 with mass $M=2.6 M_\odot$ to be a NS with a surface density compatible with nuclear saturation density with a linear EoS. In this case the model estimates its radius to be $R=10.45$ km.

In Fig. \ref{Fig:CompMR}\subref{fig:NICER}, we focus our discussion on some particular pulsars with most interest. We find that the pulsars PSR J0740+6620 (NICER+XMM) and PSR J0030+0451 (NICER) fit well with $\rho_\text{nuc}$-curve. Also we find that LIGO-Virgo constraints on the radius of canonical NS to be compatible with a surface density at $\rho_\text{nuc}$.

%%%%%%%%%%%%%%%%%%%%%%%%%%%%%%%%%%%%%%%%%%%%%% Section 6 %%%%%%%%%%%%%%%%%%%%%%%%%%%%%%%%%%%%%%%%%%%%%%%%%%%%%%%%%%%%%%%
\section{Conclusion}\label{Sec:Conclusion}
We investigated the impact of nonminimal coupling between matter and geometry, as assumed by \cite{Rastall:1972swe}, on mass-radius relation of compact objects. The theory assumes a local violation of energy conservation in presence of curved spacetime, otherwise it reduces to general relativity. Such effect should be well examined by stellar structure of compact objects whereas the spacetime is highly curvatured. The precise measurements of mass and radius of the PSR J0740+6620 by NICER would help to set a strict estimation of Rastall parameter.

For a spherically symmetric spacetime with anaisotropic matter source, we showed that the anisotropy in RT is same as in GR, which helps to clearly identify deviations from GR in terms of matter-geometry coupling. We applied KB ansatz where all physical quantities are represented in terms of Rastall parameter $\epsilon$ and the compactness parameter $C$. The precise mass-radius observational constraints from X-ray NICER+XMM observations of PSR J0740+6620 allowed us to estimate the Rastall parameter to be $\epsilon=0.041$ in the positive range. This case implies larger size for a given mass in comparison to GR which in return allows for higher compactness values to be obtained within RT framework. We showed that the additional Rastall force contributes in the hydrodynamic equilibrium to partially compensate gravitational force allowing for larger size compact star in comparison to GR for a given mass. We further showed that the upper bound on the compactness parameter is $C=0.93$ just above Buchdahl limit $C=8/9$ for isotropic case but never approaches Schwarzschild radius bound. However we utilized the dominant energy condition to constrain the maximum allowed compactness which has been found to be $C_\text{max}=0.735$, that is $2\%$ higher than GR prediction.

Interestingly, we did not assume a particular EoS, the model however in a perfect agreement with linear EoS with bag constant. More interestingly, we determined the maximum squared sound speeds at the NS core $v^2_{r}=0.332 c^2$ (radial direction) and $v^2_{t}=0.222 c^2$ (tangential direction), which satisfy the conjectured conformal bound on the sound speed $c_s^2\leq c^2/3$ at the core and also everywhere inside the NS unlike the GR case. This is unlike hadronic EoS models or Gaussian process non-parametric EoS (using NICER+XMM observations of the pulsar PSR J0740+6620) approach where the speed of sound is strongly violated $c_s^2=0.75 c^2$.

For a surface density at saturation nuclear energy $\rho_\text{nuc}=2.7\times 10^{14}$ g/cm$^3$ the model allows for a maximum mass $M=4 M_\odot$ at radius $R=16$ km which indeed gives higher compactness value in comparison to GR prediction for same surface density boundary condition. This result keeps an open window for the companion of GW190814 to be an anisotropic NS with no need to assume any exotic matter sources.

Based on the obtained results we conclude that Rastall's modified gravity and Einstein's theory of gravity are not equivalent as argued by \citet{Visser:2017gpz}. Our detailed calculations confirm that the matter-geometry coupling reconciles the sound speed within high mass compact objects $\sim 2 M_\odot$ with the conformal upper bound which is not the case in GR. This critical results among other similar work confirm the inequivalence between RT and GR as noted by \cite{Darabi:2017coc}.
%%%%%%%%%%%%%%%%%%%%%%%%%%%%%%%%%%%%%%%%%%%%%%%%%%%%%%%%%%%%%%%%%%%%
\bigskip

The author would like to thank the referee for the useful comments, which helped to improve the results.

\bibliography{j0740}
\bibliographystyle{aasjournal}

%% This command is needed to show the entire author+affiliation list when
%% the collaboration and author truncation commands are used.  It has to
%% go at the end of the manuscript.
%\allauthors

%% Include this line if you are using the \added, \replaced, \deleted
%% commands to see a summary list of all changes at the end of the article.
%\listofchanges

\end{document}